\documentclass[a4paper,usenatbib,fleqn]{aa}  

\makeatletter
\def\thickhline{%
  \noalign{\ifnum0=`}\fi\hrule \@height \thickarrayrulewidth \futurelet
   \reserved@a\@xthickhline}
\def\@xthickhline{\ifx\reserved@a\thickhline
               \vskip\doublerulesep
               \vskip-\thickarrayrulewidth
             \fi
      \ifnum0=`{\fi}}
\makeatother

\newlength{\thickarrayrulewidth}
\DeclareMathAlphabet\mathbfcal{OMS}{cmsy}{b}{n}

\usepackage{graphicx}
\usepackage{txfonts}
\usepackage{mathtools}	% Advanced maths commands
\usepackage{amssymb}	% Extra maths symbols
\usepackage{natbib}
\usepackage{caption}
\usepackage{tabularx}
\usepackage{dsfont}
\usepackage{subcaption}
\usepackage[Symbol]{upgreek}
\usepackage{epstopdf}
\usepackage{float}
\usepackage{color}
\usepackage[bookmarks=false,colorlinks,citecolor=blue,linkcolor=blue,anchorcolor=blue]{hyperref}

\DeclareMathAlphabet{\pazocal}{OMS}{zplm}{m}{n}

\newcommand{\msun}{\mathrm{M}_{\odot}}

\graphicspath{{Figures/}}

\begin{document} 

   \titlerunning{RAPTOR II}
   \title{RAPTOR II: Polarized radiative transfer in curved spacetime\thanks{The public version of {\tt RAPTOR} is available at the following URL: https://github.com/tbronzwaer/raptor}}

   \authorrunning{T. Bronzwaer et al.}
   \author{T.~Bronzwaer \inst{1},
          Z.~Younsi\inst{2,3},
          J.~Davelaar\inst{1},
          H.~Falcke\inst{1}
   }

   \institute{Department of Astrophysics/IMAPP, Radboud University Nijmegen
                  P.O. Box 9010, 6500 GL Nijmegen, The Netherlands
         \and
 Mullard Space Science Laboratory, University College London, Holmbury St.~Mary, Dorking, Surrey, RH5 6NT, United Kingdom
 \and
Institut f{\"u}r Theoretische Physik, Goethe-Universit{\"a}t Frankfurt, Max-von-Laue-Stra{\ss}e 1, D-60438 Frankfurt am Main, Germany
             }

  % \date{}

% \abstract{}{}{}{}{} 
% 5 {} token are mandatory
 
  \abstract
   { Accreting supermassive black holes are sources of polarized radiation that propagates through highly curved spacetime before reaching the observer. In order to help interpret observations of such polarized emission, accurate and efficient numerical schemes for polarized radiative transfer in curved spacetime are needed. }
   { In this manuscript we aim to extend our publicly available radiative transfer code {\tt RAPTOR} 
\citep{raptor1} to include polarized radiative transfer, so that it can produce simulated polarized observations of accreting black holes. {\tt RAPTOR} must remain compatible with arbitrary spacetimes, and it must be efficient in operation, despite the added complexity of polarized radiative transfer. }
   { We provide a brief review of different codes and methods for covariant polarized radiative transfer available in the literature and existing codes, and present an efficient new scheme. For the spacetime-propagation aspect of the computation, we develop a compact, Lorentz-invariant representation of a polarized ray. For the plasma-propagation aspect of the computation, we perform a formal analysis of the stiffness of the polarized radiative-transfer equation with respect to our explicit integrator, and develop a hybrid integration scheme that switches to an implicit integrator in case of stiffness, in order to solve the equation with optimal speed and accuracy for all possible values of the local optical/Faraday thickness of the plasma. }
   { We perform a comprehensive code verification by solving a number of well-known test problems using {\tt RAPTOR} and comparing its output to exact solutions. We also demonstrate convergence with existing polarized radiative-transfer codes in the context of complex astrophysical problems, where we found that the integrated flux densities for all Stokes parameters converged to excellent agreement.}
   { {\tt RAPTOR} is capable of performing polarized radiative transfer in arbitrary, highly curved spacetimes. This capability is crucial for interpreting polarized observations of accreting black holes, which can yield information about the magnetic-field configuration in such accretion flows. The efficient formalism implemented in {\tt RAPTOR} is computationally light and conceptually simple. The code is publicly available. 
   %\textcolor{red}{Mention speed advantage due to few coefficients: conceptually and computationally simple. Possibly mention formalism that shows d/dl p = 0.}
   }

   \keywords{radiative transfer -- black hole physics -- polarization}

   \maketitle

%-------------------------------------------------------------------

\section{Introduction}

%-description of grmhd model, disk or jet based?
%-add "normalized stokes parameter" to y axes
%-appendix numbering weird, should be A.1 etc
%-add shadow description before EHT ref
%-the Gabuzda etc papers are all about more powerful AGN and jets - just make that clear

Many low-luminosity active galactic nuclei (LLAGN) display prominent jets and compact cores that are sources of highly non-thermal continuum radio emission (see, e.g., \citealt{heeschen1970}; \citealt{wrobel1991}). The observational signatures of the compact cores have been reproduced using models that produce self-absorbed synchrotron emission in the jet (\citealt{falckebiermann}; \citealt{falckemarkoff2004}) or in a magnetized accretion flow (\citet{narayan1998ADAF}; \citet{yuan2003}; \citet{broderick2006b}; \citet{moscibrodzka2009}; \citet{dexter2009b}; see also \citet{falckenagar}). This radiation is emitted by relativistic electrons gyrating around magnetic field lines. In the optically thin limit, the emission is significantly polarized \citep{jones1979}, an effect that has been observed in higher-luminosity AGN sources (\citealt{gabuzda1996}; \citealt{gabuzda2000}; \citealt{gabuzda2005}). The polarized emission from an accreting AGN can therefore yield information about the source's magnetic-field morphology, which may be crucial to the evolution of the AGN's accretion flow. The Event Horizon Telescope (EHT) is a worldwide millimeter-wavelength array capable of resolving the black-hole shadow (\citealt{goddi2016}; \citealt{EHT1}), a characteristic feature of the radio-frequency emission from optically thin AGN at the scale of the event horizon (\citealt{falcke2000}; \citealt{brodericknarayan2006}), although it may be obscured or exaggerated in certain accretion scenarios (see \citet{gralla2019} and \citet{narayan2019}). The EHT can also determine the polarization state of such emission: \cite{johnson2015} reports 1.3-mm observations (230 GHz) that indicate partially ordered magnetic fields within a region of about 6 Schwarzschild radii around the event horizon of Sagittarius A* (Sgr A*), the supermassive black hole in the center of the Milky Way. \cite{bower2003} report stable long-term behavior and short-term variability in Sgr A* rotation measure, implying a complex inner region (within 10 Schwarzschild radii) in which both emission and propagation effects are important to the observed polarization. \cite{hada2016} study the central black hole in the galaxy M87, observing a bright feature with (linear) polarization degree of 0.2 at 86 GHz at the jet base. Observations in infrared by \cite{gravity2018} were consistent with a model in which a relativistic `hot spot' of material, orbiting near Sgr A*'s innermost stable circular orbit (ISCO) in a poloidal magnetic field, emits polarized synchrotron radiation.

To study accreting supermassive black holes, general-relativistic radiative-transfer (GRRT) codes are used (see, e.g., \citealt{jaroszynski1997}; \citealt{bromley2001}; \citealt{broderick2006}; \citealt{noble2007}; \citealt{dexter2009}; \citealt{younsi2012}). GRRT codes solve the geodesic equation in curved spacetime to compute null geodesics, and then solve the radiative-transfer equation along the null geodesics to produce an image. Doing so requires the evaluation of emission and absorption coefficient of radiation along the geodesics. In our case, the emission and absorption coefficients are computed using the state variables of a radiating plasma (the black-hole accretion flow, generally consisting of both a disk and a jet). The plasma variables (such as density, magnetic fields, and temperature) needed to compute the emission and absorption coefficients are either provided by analytical or semi-analytical models, or by fully numerical, general-relativistic magnetohydrodynamical (GRMHD) plasma simulations. Some models, such as the thin-disk model of an accreting black hole in a black-hole binary system (\citet{shakura1973} and \citet{novikovthorne1973}), consist of a geometrically thin, optically thick disk, meaning that the emissivity function along a ray is a delta function; other models, such as those based on GRMHD data, may be geometrically thick yet optically thin, necessitating the use of volumetric rendering techniques, specifically, solving the radiative-transfer equation along a ray.

In order to interpret and complement polarized observations, it is important that the numerical radiative-transfer tools used to study accreting supermassive black holes are capable of including polarization. In a previous paper \citep{raptor1}, we presented {\tt RAPTOR}, a publicly available GRRT code capable of performing time-dependent radiative transfer in arbitrary spacetimes. In the present paper, we develop a novel formalism and algorithm for polarized radiative transfer in order to extend {\tt RAPTOR}'s capabilities. Two things affect the polarization state of radiation propagating through a plasma in a strong gravitational field, namely: $i)$ propagation through curved spacetime itself (which may rotate the polarization vector around the ray's propagation direction), and $ii)$ interaction with the plasma (which may affect the polarization state in a general way). In order to model these two processes quantitatively, several equivalent formalisms for covariant polarized radiative transfer have been proposed in the literature; they differ in the ways in which they represent polarized radiation, as well as the method of integration. \citet{broderick2004}, \citet{shcherbakov2011}, \citet{dexter2016}, and \citet{younsi2020} represent polarized radiation using the Stokes parameters plus a polarization basis vector, integrating the Stokes parameters through curved spacetime and any plasma that may be present while parallel-transporting the basis vector. \citet{ipole} also represents polarized radiation using the Stokes parameters for computing the plasma interaction, but their formalism, which is based on \citet{gammie2012}, transforms back and forth between the Stokes parameters (which are convenient for computing the ray's interaction with a radiating plasma) and a covariant description of the polarization state, a tensor called the coherence matrix, which is convenient for propagation through spacetime. In this work, we choose to develop a novel formalism for {\tt RAPTOR}, to match with our previously chosen numerical methods \citep{raptor1} and to optimize {\tt RAPTOR}'s computational efficiency and accuracy.

This paper is organized as follows: in Section \ref{sec:polarized_rad_trans}, we discuss the theory of polarized radiative transfer in curved spacetime, as well as different methods for solving the governing equations. We also present {\tt RAPTOR}'s representation of polarized radiation. In Section \ref{sec:implementation}, we construct a numerical algorithm that solves the polarized radiative-transfer equation and analyze that equation's stiffness with respect to our integrator, using the results to optimize our algorithm's accuracy. We demonstrate the correctness of our algorithm by comparing {\tt RAPTOR} output to previous results, as well as the output of other codes, in the context of complex, astrophysical problems in Section \ref{sec:verification}. We summarize our results in Section \ref{sec:discussion}.

%\textcolor{red}{Redo images 4 x 4 plots, higher res, and comment on them more in text}

%\textcolor{red}{Additional inclinations for polarized result, 20, 60, 90 choose mass minimum/maximum. How different is the image structure? Write paragraph on how polarization data will help us determine inclination.}

\section{Polarized Radiative Transfer In Curved Spacetime}
\label{sec:polarized_rad_trans}
%\ZY{Best to start with a brief overview of the polarised radiation transport equation in astrophysics, it's importance etc., then mention the first Raptor paper.}

Electromagnetic radiation is the most ubiquitous messenger of information in astrophysics. Emitted by sources widely distributed in space and time, it pervades the universe and interacts with matter, most of which exists as a plasma. Some of this radiation is emitted in a highly polarized state, such as synchrotron radiation. Some of it is (de)polarized after emission, for example, by interaction with magnetized plasmas \citep{aitken2000} or dust grains \citep{davis1951}. Besides interaction with matter, propagation of polarized radiation through curved spacetime can also affect the radiation's polarization state \citep{MTW}.

In \citet{raptor1}, we represented an `unpolarized' ray of light by its intensity $I_{\nu}$ (where $\nu$ is the radiation frequency), position $x^{\mu}$, and wave vector $k^{\mu}$, the wave vector describing the ray's traveling direction and frequency, in order to solve the radiative-transfer equation along null geodesics. In the case of polarized radiative transfer, one must additionally keep track of the ray's polarization state, which describes the orientation and phase of the electromagnetic oscillations associated with the ray. In the case of an ensemble of photons with identical polarization states, the polarization state is said to be pure, while in the case of an ensemble of photons with different polarization states, the polarization state is mixed. As before, {\tt RAPTOR} functions in the regime of geometrical optics, in which the radiation's wavelength must be much smaller than the typical length scales of plasma features and spacetime curvature.

\subsection{Descriptions of a polarized ray; propagation through a curved spacetime}

Various descriptions of the polarization state of radiation are used in the literature. These descriptions differ in three key ways: $i)$ whether or not they encode the phase of the polarization state; $ii)$ whether or not they can describe mixed states or only pure states; and $iii)$ whether or not they are Lorentz covariant.

Note that in the present context, we neglect all information about the phase of the polarization state. This is because ray tracing is valid only in the regime of geometrical optics, in which the phase is omitted, so that wave effects (such as interference and diffraction) are neglected. We do, however, choose to incorporate a description of mixed states, as astrophysical sources can emit polarized radiation in such states. Additionally, a particular description may be more suitable than others, depending on the circumstances. For example, interaction with a radiating plasma is commonly described using the Stokes parameters, while the effects of propagation through curved spacetime are more easily calculated using a Lorentz-covariant description. It is therefore necessary to be able to convert between the different descriptions. In this section we describe the two descriptions of a polarized ray used in {\tt RAPTOR}, as well as the transformation equations between them. We then present the equations of motion through curved spacetime for a polarized ray.

\subsubsection{Stokes parameters}
\label{sec:stokes}

The Stokes parameters, denoted $I, Q, U, V$, describe a ray's polarization state by encoding the total intensity of radiation, the intensities of the two types of linear polarization, and the circularly polarized intensity, respectively. The Stokes parameters must be defined in a particular coordinate system such that the wave vector of the associated radiation is parallel to the coordinate system's $z$-axis. Their signs are then determined by choosing a convention, i.e., a handedness and orientation of the observer coordinate frame. This paper follows the IAU/IEEE convention, in which the angle $\chi = 1/2 \arctan{\left(U/Q\right)}$, called the electric vector position angle (EVPA), is defined as an angle East of North, and the sense of circular polarization is called right-handed (left handed) if the direction of rotation of the EVPA is clockwise (anti-clockwise) for an observer looking in the direction of propagation. For a detailed description of this convention, see, for example, \citet{hamaker1996}.

 The Stokes parameters are often represented as a vector, the Stokes vector, denoted $\text{\bf S}=(I,Q,U,V)^{\rm T}$. They may also be written in terms of specific intensities, which is convenient for our purpose: $\text{\bf S}_{\nu}=(I_{\nu},Q_{\nu},U_{\nu},V_{\nu})^{\rm T}$. The Stokes parameters represent quantities that can be readily measured, hence they are generally used to report observational results. 

The Stokes parameters can encode both pure and mixed polarization states; one can distinguish between the two using the degree of polarization, $p$, which is calculated as follows:
\begin{equation}
p = \frac{I_{\nu,\text{pol}}}{I_{\nu}} = \frac{\sqrt{Q_{\nu}^2 + U_{\nu}^2 + V_{\nu}^2}}{I_{\nu}},
\label{eqn:deg_of_p}
\end{equation}
where $I_{\nu,\text{pol}}$ is the intensity of polarized radiation (which is in a pure state, described by $Q_{\nu},U_{\nu},V_{\nu}$) and $I_{\nu}$ the total intensity.
For pure polarization states, this results in $p=1$ and $I_{\nu,\text{pol}} = I_{\nu}$; for mixed states, we have $0 \leq p < 1$ and $I_{\nu,\text{pol}} \leq I_{\nu}$.
The Stokes parameters omit phase information, and the Stokes vector is not Lorentz covariant. 
% rotating the coordinate system about the wave vector changes the orientation of the x and y axis, and thus the values of the Stokes parameters.
Consequently, in order to transport the Stokes parameters through curved spacetime consistently, it is necessary to transport a basis vector along the geodesic, even in the case of propagation through a vacuum. \citet{shcherbakov2011} present an algorithm that integrates the Stokes parameters directly through curved spacetime in this manner.

Just as the Lorentz-invariant quantity $\mathcal{I} = I_{\nu} / \nu^3$ was employed during integration of the radiative-transfer equation in \citet{raptor1}, Lorentz-invariant Stokes intensities are defined as follows:
\begin{equation}
\mathbfcal{S} := \frac{\text{\bf S}_{\nu}}{\nu^3},
\end{equation}
where $\nu$ represents the ray's frequency in the frame in which $\text{\bf S}_{\nu}$ is evaluated. It is convenient to use these Lorentz-invariant quantities during integration, when constantly shifting between frames.

\subsubsection{Polarization four-vector}

The polarization four-vector, $f^{\mu}$, is a complex-valued vector that describes a pure polarization state. As it is a four-vector, it is Lorentz covariant. The polarization four-vector is a unit vector:
\begin{equation}
    f^{\mu} f_{\mu}^* = 1,
\end{equation}
where the asterisk denotes complex conjugation. The polarization vector associated with a ray is parallel-transported along the ray's null geodesic:
\begin{equation}
    k^{\alpha} \nabla_{\alpha} f^{\mu} = 0.
    \label{eqn:f_transport}
\end{equation}
When expressed in a suitable tetrad frame (see Section~\ref{sec:fluidframe}), and provided the frame is inertial, meaning that its acceleration vector vanishes, the polarization vector's components represent normalized, projected electric-field amplitudes along the frame's $x$ and $y$-axes, respectively. Using Roman indices framed by parentheses to indicate tetrad-frame coordinates, we have:
\begin{equation}
f^{\left( a \right)} = e^{\left( a \right)}_{\mu} f^{\mu} = \begin{pmatrix}
           0 \\
           \hat{E}_x \\
           \hat{E}_y \\
           0
         \end{pmatrix},
\end{equation}
where $e^{\left( a \right)}_{\mu}$ are the components of the $a$-th tetrad-basis vector expressed in coordinates $\mu$, and $\hat{E}_x$ and $\hat{E}_y$ are components of a unit vector pointing along the electric field. In the most general case, both $\hat{E}_x$ and $\hat{E}_y$ are complex, and the polarization vector encodes the polarization state's overall phase. It is also possible to restrict one of the components to be real; only the overall phase information is then lost.

The polarization four-vector only describes pure polarization states; it cannot describe mixed states. However, by keeping track of both the intensity of polarized radiation, $\mathcal{I}_{\text{pol}}$ and the total intensity $\mathcal{I}$ in addition to $f^{\mu}$, it is possible to represent rays with a mixed polarization state. 

Given the triplet $\left( \mathcal{I}, \mathcal{I}_{\text{pol}}, f^{\mu} \right)$, plus a suitable tetrad in which to express the Stokes parameters, the transformation $ \left( \mathcal{I}, \mathcal{I}_{\text{pol}}, f^{\left( a \right)} \right) \rightarrow{} \mathbfcal{S}$ is given by
\begin{equation}
\mathbfcal{S} = \begin{pmatrix}
           \mathcal{I} \\
           \mathcal{Q} \\
           \mathcal{U} \\
           \mathcal{V}
         \end{pmatrix} = 
         \begin{pmatrix}
           \mathcal{I} \\
           %E_x E_x^* - E_y E_y^* \\
           \mathcal{I}_{\text{pol}} \left( f^{\left( 1 \right)} f^{\left( 1 \right)*} - f^{\left( 2 \right)} f^{\left( 2 \right)*} \right) \\
           %E_x E_y^* + E_y E_x^* \\
           \mathcal{I}_{\text{pol}} \left( f^{\left( 1 \right)} f^{\left( 2 \right)*} + f^{\left( 2 \right)} f^{\left( 1 \right)*} \right) \\
           %i \left( E_x E_y^* - E_y E_x^* \right)
           i \mathcal{I}_{\text{pol}} \left( f^{\left( 1 \right)} f^{\left( 2 \right)*} - f^{\left( 2 \right)} f^{\left( 1 \right)*} \right)
         \end{pmatrix}. 
         \label{eqn:f_to_Stokes}
\end{equation}
%The case $I_{\text{pol}}=0$ implies no polarized radiation. In this case, {\tt RAPTOR} will not have created a polarization vector in the first place, meaning that $I_{\text{pol}} > 0$ whenever Eq.~\ref{eqn:f_to_Stokes} is evaluated. 
%\textcolor{red}{Be consistent with using column/row vectors; Stokes vector is now presented in both forms without clarification}
The reverse transformation, $\mathbfcal{S} \rightarrow{} \left( \mathcal{I}, \mathcal{I}_{\text{pol}}, f^{\left( a \right)} \right)$, is degenerate, as the latter quantity contains an additional degree of freedom (the overall phase of the ray's polarization state). The degeneracy is lifted by choosing a phase, i.e., by demanding that $f^{\left( 1 \right)} \in \mathbb{R}$. $f^{\left( a \right)}$ is then computed as follows: 
\begin{subequations}
\begin{align}
  \mathcal{I_{\rm pol}} &= \sqrt{\mathcal{Q}^2 + \mathcal{U}^2 + \mathcal{V}^2}, \\
  f^{\left( 1 \right)} &= \sqrt{\frac{1 + \tilde{Q}}{2}}, \\
  f^{\left( 2 \right)} &=
  \begin{cases}
    1 & \text{if $f^{\left( 1 \right)} = 0$}, \\
    \cfrac{\tilde{U} - i \tilde{V}}{2 f^{\left( 1 \right)} } & \text{otherwise},
  \end{cases} 
\end{align}
\end{subequations}
where $\tilde{Q} \equiv \mathcal{Q} / \mathcal{I}_{\text{pol}}$, and similarly for $\tilde{U}$ and $\tilde{V}$ (note that $\mathcal{I}$ retains its identity when transforming).

\subsubsection{Spacetime propagation equation}

In our model, the polarization state of a ray is affected by two processes: propagation through spacetime and interaction with a plasma. Given Eqs.~\ref{eqn:deg_of_p} and \ref{eqn:f_transport}, we can express the equations of motion for propagation of a polarized ray through curved (vacuum) spacetime:
\begin{subequations}
\begin{align}
    &\frac{{\rm d}}{{\rm d}\lambda} \Biggr\rvert_{S} f^{\mu} = - \Gamma^{\mu}_{\ \alpha \rho} k^{\alpha}f^{\rho}, \label{eqn:spacetime_propagation} \\
    &\frac{{\rm d}}{{\rm d}\lambda} \Biggr\rvert_{S} \mathcal{I} = 0, \\ 
    &\frac{{\rm d}}{{\rm d}\lambda} \Biggr\rvert_{S} \mathcal{I}_{\text{pol}} = 0, \label{eqn:Ipol_propagation}
\end{align}
\end{subequations}
where $\lambda$ is the affine parameter, expressed in units of length ($GM/c^2$), which parametrizes the null geodesic. The subscript $S$ implies that we only consider effects due to propagation through curved spacetime, ignoring the plasma.
Equations~\ref{eqn:spacetime_propagation}-\ref{eqn:Ipol_propagation} show that the degree of polarization $p$ remains constant along a ray when propagated through a curved (vacuum) spacetime. This is a consequence of the fact that $I_{\nu,\text{pol}}$ and $I_{\nu}$ transform in the same way between different frames, so that their ratio remains a constant from frame to frame, and thus throughout a (vacuum) spacetime integration. Equivalently, the corresponding Lorentz-invariant intensities $\mathcal{I}$ and $\mathcal{I}_{\text{pol}}$ are themselves constant along the ray, as is the ratio between them.

\subsection{Constructing suitable tetrad frames in which to express the Stokes parameters}
\label{sec:fluidframe}

%\textcolor{red}{MOVE THIS SECTION ELSEWHERE? PERHAPS IMPLEMENTATION, OR AFTER PLASMA INTERACTION.
%SO FAR, EVERYTHING DONE IN COORD FRAME. NEXT, IN PLASMA FRAME. THUS NEED A WAY TO TRANSFORM BETWEEN THEM...}
As in the case of unpolarized radiative transfer, it will be convenient to perform the radiative-transfer computations in a suitably chosen frame. Unlike in the case of unpolarized radiative transfer (where transforming between frames is accomplished simply by computing a ray's frequency seen by an observer co-moving with the frame of interest), this frame must now be constructed explicitly, as polarized radiative-transfer computations depend on the orientation of the frame. This must be done both at the observer's location (the camera) and also at any location where the ray interacts with radiating matter. It is also necessary to specify the handedness of the tetrad frame, which is achieved by ordering its basis vectors. 

We employ a generalized version of the tetrad described in \cite{gammie2012}; as is the case for those authors, our tetrad-frame indices $\left(t, \parallel, \perp, K\right)$ correspond with $\left(t, x, y, z\right)$, respectively, defining a right-handed coordinate system. As previously, we adopt the $\left(-,+,+,+\right)$ metric convention, and we denote the tetrad-frame coordinates with Roman indices enclosed in parentheses. In what follows, it is assumed that the ray's wave vector is null ($k_{\alpha} k^{\alpha} = 0$), that the frame's velocity four-vector is timelike with norm $-1$, i.e., $u_{\alpha} u^{\alpha} = -1$, and that $d^{\alpha}$ is a spacelike vector ($d_{\alpha} d^{\alpha} > 0$). When the ray resides inside the volume for which GRMHD data is available, $u^{\alpha}$ is given by the local plasma four-velocity, and we may use the local magnetic-field four vector ($b^{\alpha}$) for the spacelike vector $d^{\alpha}$; $b^{\alpha}$ is spacelike (except for the pathological case in which it vanishes, in which case the integration step may be skipped), and orienting the tetrad this way will simplify calculations because all coefficients related to Stokes U vanish. Under these circumstances we shall refer to the tetrad frame as the plasma frame. When no magnetic-field vector is available, or when we wish to orient the tetrad in a specific way (e.g., in order to represent a particular observer's position and attitude), the trial vector $d^{\alpha}$ must be constructed in such a way that it is spacelike. This generally happens at the observer's location, in which case we shall refer to the tetrad as the observer frame; the velocity four-vector in such cases is generally chosen to be stationary, so that $u_{\mu{\rm ,obs}} = \left( u_t, 0, 0, 0 \right)$. Since $u^t$ is a constant, the observer frame is an inertial frame. Note that the tetrad must be constructed so as to respect the IAU/IEEE observer convention \citep{hamaker1996}. 

To construct a tetrad frame for a ray, we start by defining the intermediate quantities
\begin{subequations}
\begin{align}
d^2 &\coloneqq d_{\alpha} d^{\alpha}, \\
\beta &\coloneqq u_{\alpha} d^{\alpha}, \\
\omega &\coloneqq -k_{\alpha} u^{\alpha}, \\
\mathcal{C} &\coloneqq \frac{k_{\alpha} d^{\alpha}}{g} - \beta, \\
\mathcal{N} &\coloneqq \sqrt{d^2 + \beta^2 -\mathcal{C}^2}.
\end{align}
\end{subequations}
The tetrad is then constructed as follows, using the Gram-Schmidt orthonormalization procedure:
\begin{subequations}
\begin{align}
e^{\mu}_{(t)} &= u^{\mu} \ , \\
e^{\mu}_{(K)} &= \frac{k^{\mu}}{\omega} - u^{\mu} \ , \\
e^{\mu}_{(\parallel)} &= \frac{d^{\mu} + \beta u^{\mu} - \mathcal{C} e^{\mu}_{(K)}}{\mathcal{N}} \ , \\
e^{\mu}_{(\perp)} &= \frac{\epsilon^{\mu\nu\alpha\beta}u_{\nu}k_{\alpha}d_{\beta}}{\omega \mathcal{N}} \ ,
\end{align}
\end{subequations}
where 
\begin{equation}
\epsilon^{\mu\nu\alpha\beta} = -\frac{1}{\sqrt{-g}}\left[ \mu\nu\alpha\beta \right] \ ,
\end{equation}
is the Levi-Civita tensor, $\left[ \mu\nu\alpha\beta \right]$ is the permutation symbol, and $g\equiv \mathrm{det}\left(g_{\mu\nu}\right)$. Note that the general metric tensor $g_{\mu \nu}$ acts on coordinate-frame indices while the Minkowski metric tensor $\eta_{\mu \nu}$ acts on fluid-frame indices, e.g., $g_{\mu \nu} \to \eta_{\mu \nu}$ in the fluid frame.

%Note that we follow the IEEE convention for polarization, meaning that the sense of polarization is called right handed (left handed) if the direction of rotation is clockwise (anti-clockwise) for an observer looking in the direction of propagation. \textcolor{red}{Update the preceding to be in accordance with EHT, which doesn't use the IEEE convention. Possibly set a switch in RAPTOR for IEEE or EHT standards)}

\subsection{Plasma interaction; synchrotron emission, absorption, and Faraday rotation coefficients}
\label{sec:coefficients}

The ray's interaction with a plasma is most conveniently expressed using the Stokes parameters. Given the plasma frame, we require a set of emission, absorption, and rotation coefficients $j$, $\alpha$, and $\rho$, respectively. The coefficients used in {\tt RAPTOR} are adapted from \citet{dexter2016}. They are recapitulated in Appendix \ref{app:coefficients}. Note that the coefficients must be expressed in their Lorentz-invariant form (Section \ref{sec:stokes}). The effect of the plasma on the invariant Stokes parameters is given by
\begin{equation}
\frac{{\rm d}}{{\rm d}\lambda} \Biggr\rvert_{P} \begin{pmatrix}
           \mathcal{I} \\
           \mathcal{Q} \\
           \mathcal{U} \\
           \mathcal{V}
         \end{pmatrix} = 
         \begin{pmatrix}
           j_I \\
           j_Q \\
           j_U \\
           j_V
         \end{pmatrix} -
         \begin{pmatrix}
           \alpha_I & \alpha_Q & \alpha_U & \alpha_V \\
           \alpha_Q & \alpha_I & \rho_V   & -\rho_U  \\
           \alpha_U & -\rho_V  & \alpha_I & \rho_Q   \\
           \alpha_V & \rho_U   & -\rho_Q  & \alpha_I
         \end{pmatrix}
         \begin{pmatrix}
           \mathcal{I} \\
           \mathcal{Q} \\
           \mathcal{U} \\
           \mathcal{V}
         \end{pmatrix},
\label{eqn:plasma_interaction}
\end{equation}
with the subscript $P$ implying that only the ray's interaction with the plasma is taken into account (ignoring the effects due to spacetime propagation). 
%\ZY{Recapitulate these in the appendix, with typos corrected from the original papers, for the readers' convenience?}

\section{Implementation}
\label{sec:implementation}

In this section, we develop algorithms in order to implement the polarized radiative-transfer formalism described in Section \ref{sec:polarized_rad_trans} in {\tt RAPTOR}. The main challenge in this implementation lies in the fact that the polarized radiative-transfer equation, Eq.~\ref{eqn:plasma_interaction}, may become stiff with respect to explicit integrators depending on the plasma conditions. To mitigate this problem, we analyze when stiffness occurs, and develop an implicit integrator for such integration steps. Precise knowledge of where stiffness occurs is crucial when it comes to minimizing the number of implicit steps, which are less accurate, although much more stable.

\subsection{Integration strategy}

As in the case of unpolarized radiative transfer through curved spacetime, the integration can be thought of as consisting of two parts: vacuum integration (which takes care of the effects on a ray's polarization state due purely to traveling through curved spacetime) and plasma integration (which describes the ray's interaction with the plasma through the plasma's emission, absorption, and rotation coefficients (Section \ref{sec:coefficients})). The two sub-problems are handled with separate routines. During an integration step, and when the ray resides in a radiating plasma, the plasma interaction is computed first, after which a spacetime-propagation step is taken (as if through a vacuum). When the ray resides in vacuum, the plasma-integration step is omitted.

\subsection{Numerical scheme for plasma integration}
\label{sec:plasma_integration}

In order to take a ray's interaction with the radiating plasma into account, we must solve Eq.~\ref{eqn:plasma_interaction} numerically. Note that, due to aligning the frame so that the Stokes U polarization mode is parallel to the plasma's magnetic-field vector, we have $j_U=\alpha_U=\rho_U=0$. Depending on the local values of the emission, absorption, and rotation coefficients, Eq.~\ref{eqn:plasma_interaction} may be a stiff equation for explicit integration schemes. Note that the condition for stiffness is different for each explicit integrator, and that it does not directly depend on any particular physical quantity, but rather on the product of the local step size taken by {\tt RAPTOR} and the largest eigenvalue of the matrix appearing in Eq.~\ref{eqn:plasma_interaction} (see, e.g., \citet{gautschi2011}). {\tt RAPTOR} uses the fourth-order Runge-Kutta (RK4) explicit integrator (see below); an analysis of when Eq.~\ref{eqn:plasma_interaction} becomes stiff for the explicit RK4 integrator is presented in Appendix \ref{sec:stiffness}. The result of that analysis is a numerical stiffness check that is performed at each integration step.
 
Stiff equations are practically impossible to solve using forward integration methods of the order considered so far in {\tt RAPTOR}; the solution will become unstable, and prohibitively small step sizes are required. Implicit integration methods are much more stable, and can be applied even to stiff problems. However, they tend to dampen rapid oscillations, which increases the stability, but which comes at the cost of less accuracy. In order to integrate Eq.~\ref{eqn:plasma_interaction}, \cite{dexter2016} employs three integration strategies: an implicit-explicit integrator routine from the {\tt LSODA} package, an implementation of the DELO method \citep{rees1989}, and an explicit quadrature method based on the work of \citet{LDI2}. \citet{moscibrodzka2016} employs a semi-analytical solution also based on \citet{LDI2}, as well as a three-step numerical integration routine. We choose to develop a novel implicit-explicit integrator for {\tt RAPTOR}. Since stiffness conditions vary throughout the plasma, using only implicit methods may needlessly sacrifice accuracy in regions where Eq.~\ref{eqn:plasma_interaction} is not stiff. Switching to an explicit integration scheme in such regions improves the overall accuracy of the computation.

For explicit integration steps, our algorithm uses the RK4 integrator:
\begin{subequations}
\begin{align}
\text{\bf C}_{1,\mathbfcal{S}} &= \Delta \lambda \cdot \text{{\bf F}} \Big( \mathbfcal{S} \Big), \\
\text{\bf C}_{2,\mathbfcal{S}} &= \Delta \lambda \cdot \text{{\bf F}} \left( \mathbfcal{S} + \frac{1}{2} \text{\bf C}_{1,\mathbfcal{S}} \right), \\
\text{\bf C}_{3,\mathbfcal{S}} &= \Delta \lambda \cdot \text{{\bf F}} \left( \mathbfcal{S} + \frac{1}{2} \text{\bf C}_{2,\mathbfcal{S}} \right), \\
\text{\bf C}_{4,\mathbfcal{S}} &= \Delta \lambda \cdot \text{{\bf F}} \Big( \mathbfcal{S} + \text{\bf C}_{3,\mathbfcal{S}} \Big),
\end{align}
\label{eqn:RK4_plasma}%
\end{subequations}
where $\text{{\bf F}}$ represents the right-hand-side of Eq.~\ref{eqn:plasma_interaction}. A single explicit integration step proceeds as follows:
\begin{equation}
    \mathbfcal{S}_{new} = \mathbfcal{S} + \frac{1}{6} \Big( \text{\bf C}_{1,\mathbfcal{S}} + 2 \text{\bf C}_{2,\mathbfcal{S}} + 2 \text{\bf C}_{3,\mathbfcal{S}} + \text{\bf C}_{4,\mathbfcal{S}}\Big).
\end{equation}
For implicit steps, our algorithm uses the (second-order) implicit trapezoid method:
\begin{equation}
\mathbfcal{S}_{new} = \mathbfcal{S} + \frac{ \Delta \lambda}{2} \Bigg[ \Big( \text{\bf j} - \text{\bf M} \mathbfcal{S}_{new} \Big) + \Big( \text{\bf j} - \text{\bf M} \mathbfcal{S} \Big) \Bigg],
\label{eqn:implicit_euler}
\end{equation}
where $\text{\bf j} = \left( j_I, j_Q, j_U, j_V \right)^{\rm T}$ and $\text{\bf M}$ is the 4-by-4 matrix appearing in Eq.~\ref{eqn:plasma_interaction}. Since Eq.~\ref{eqn:plasma_interaction} is linear, the implicit trapezoid method for this equation yields an explicit equation for $\mathbfcal{S}_{new}$ using an LU-decomposition, so that there is no root-finding penalty even for implicit steps (see Appendix \ref{app:implicit_euler}).

%For the switching criteria, \textcolor{red}{we use...}
%\ZY{It would also be great to show some demos of what images or particular geodesics look like without and with these switching criteria activated.}

\subsection{Numerical scheme for vacuum integration}

Previously we integrated a ray's position, $x^{\alpha}$, as well as its wave vector, $k^{\alpha}$, through arbitrary, curved spacetimes. We must now also keep track of the description of the ray's polarization state, which is captured in the polarization vector $f^{\mu}$; in other words, we must solve Eq.~\ref{eqn:spacetime_propagation}.

As before, we use a fourth-order Runge-Kutta scheme to integrate the ray backward, i.e., starting at the camera. After a stopping condition has been reached (e.g., the ray plunges into the horizon, or it reaches the outer boundary of a GRMHD simulation), polarized radiative transfer proceeds in the forward direction, i.e., toward the camera. We extend Eqs.~7-14 from \citet{raptor1} (noting that forward integration, i.e., from the plasma toward the camera, is employed) to include the polarization vector as follows: 
\begin{subequations}
\begin{align}
C_{1,f^{\alpha}} &=& \Delta \lambda \cdot F^{\alpha}\left(x^{\underline{i}},k^{\underline{i}},f^{\underline{i}} \right), \\
C_{2,f^{\alpha}} &=& \Delta \lambda \cdot F^{\alpha}\left(x^{\underline{i}} + \tfrac{1}{2} C_{1,x^{\underline{i}}}, k^{\underline{i}} +
\tfrac{1}{2} C_{1,k^{\underline{i}}}, f^{\underline{i}} + \tfrac{1}{2} C_{1,f^{\underline{i}}}  \right), \\
C_{3,f^{\alpha}} &=& \Delta \lambda \cdot F^{\alpha}\left(x^{\underline{i}} + \tfrac{1}{2} C_{2,x^{\underline{i}}}, k^{\underline{i}} +
\tfrac{1}{2} C_{2,k^{\underline{i}}}, f^{\underline{i}} + \tfrac{1}{2} C_{2,f^{\underline{i}}} \right), \\
C_{4,f^{\alpha}} &=& \Delta \lambda \cdot F^{\alpha}\left(x^{\underline{i}}+ C_{3,x^{\underline{i}}}, k^{\underline{i}} + C_{3,k^{\underline{i}}}, f^{\underline{i}} + C_{3,f^{\underline{i}}} \right),
\end{align}
\label{eqn:RK4_geodesic}%
\end{subequations}
where, as in \citet{raptor1}, underlined indices are a notational shorthand to indicate that all coordinate indices occur in the right-hand side of these equations, and $F^{\alpha}$ represents the right-hand side of Eq.~\ref{eqn:spacetime_propagation}. Given these coefficients, an integration step proceeds as follows:
\begin{equation}
    f^{\alpha}_{new} = f^{\alpha} + \frac{1}{6} \left( C_{1,f^{\alpha}} + 2 C_{2,f^{\alpha}} + 2 C_{3,f^{\alpha}} + C_{4,f^{\alpha}}\right).
\end{equation}

\section{Code Verification}
\label{sec:verification}

In this section we aim to verify the correctness of {\tt RAPTOR} output by comparing it to analytical results as well as output from different codes. For this purpose, a number of verification tests were selected from the literature and reproduced using {\tt RAPTOR}. Convergence tests were also performed for all integrators described in the previous section. 

\subsection{Plasma-integration test; comparison to analytical solution}
\label{sec:plasmatests}

As a first step, we test our numerical integrator for Eq.~\ref{eqn:plasma_interaction}, i.e., the ray's interaction with the radiating plasma. Note that the two tests reproduced in this section were also reproduced by \citet{dexter2016} and \citet{ipole}, the latter of which performed the test using non-standard `snake' coordinates, which means that the space-time integration routine is tested simultaneously in their case, making it a more challenging test. In each case, the initial conditions are $I=Q=U=V=0$ and the stepsize is $\Delta s = 0.003$.

\begin{figure}
\centering
\includegraphics[width=0.5 \textwidth]{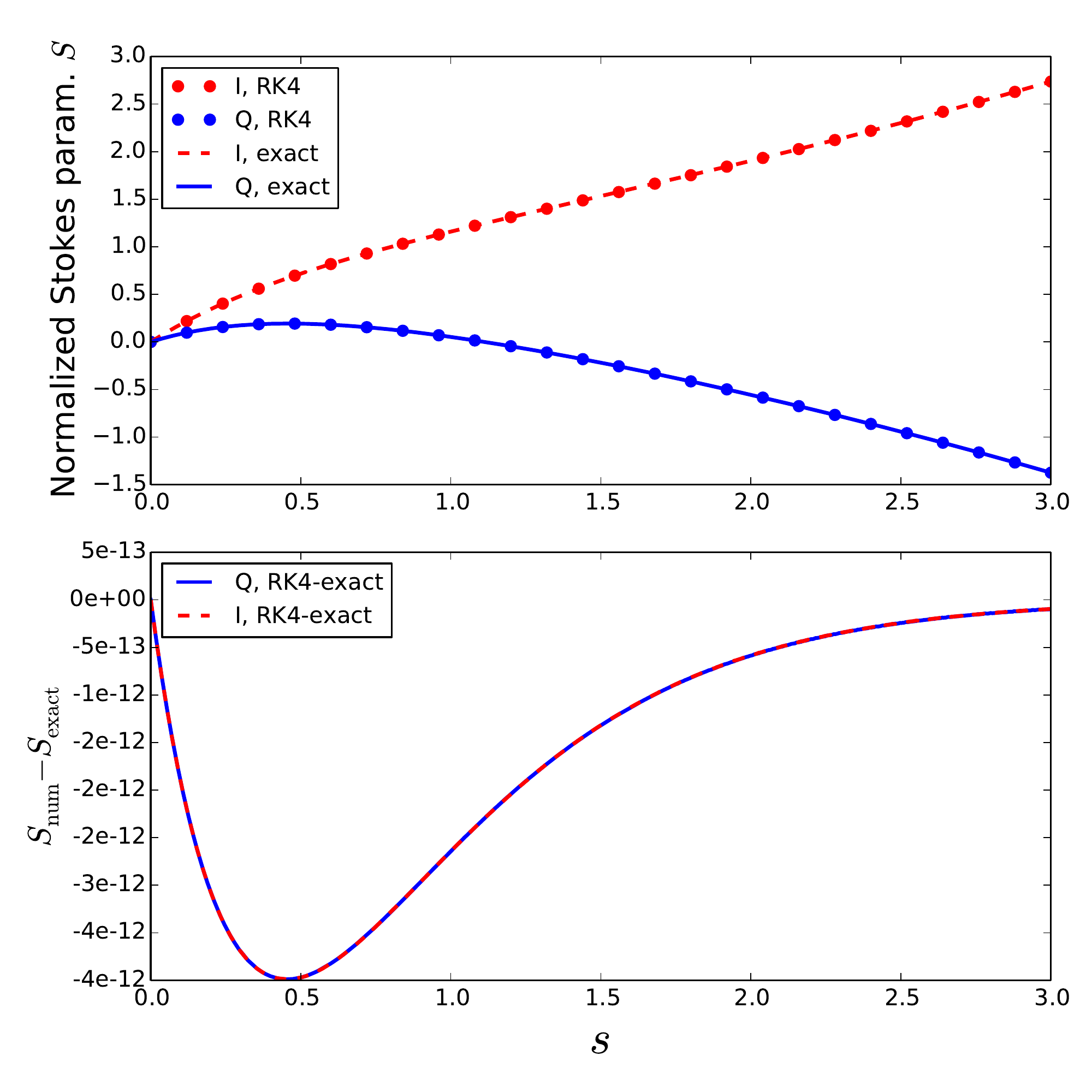}
\caption{A plot of Stokes I and Q (in normalized units) as a function of distance traveled $s$, using the explicit Runge-Kutta integration routine, for the first flat-spacetime plasma-integration test.}\label{fig:plasmatest1}
\end{figure}

\begin{figure}
\centering
\includegraphics[width=0.5 \textwidth]{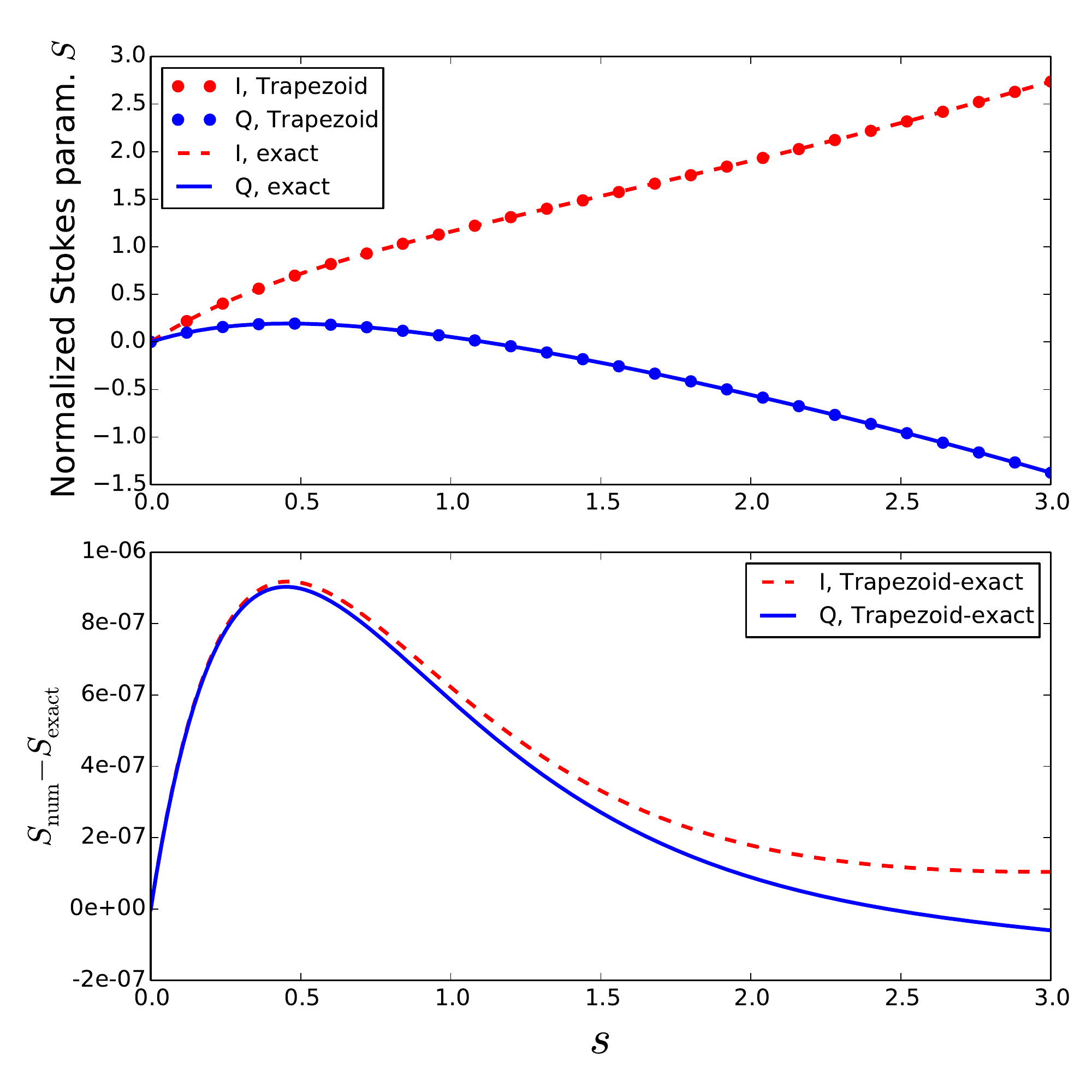}
\caption{A plot of Stokes I and Q (in normalized units) as a function of distance traveled $s$, using the implicit trapezoid integration routine, for the first flat-spacetime plasma-integration test.}\label{fig:plasmatest1_IE}
\end{figure}

\begin{figure}
\centering
\includegraphics[width=0.5 \textwidth]{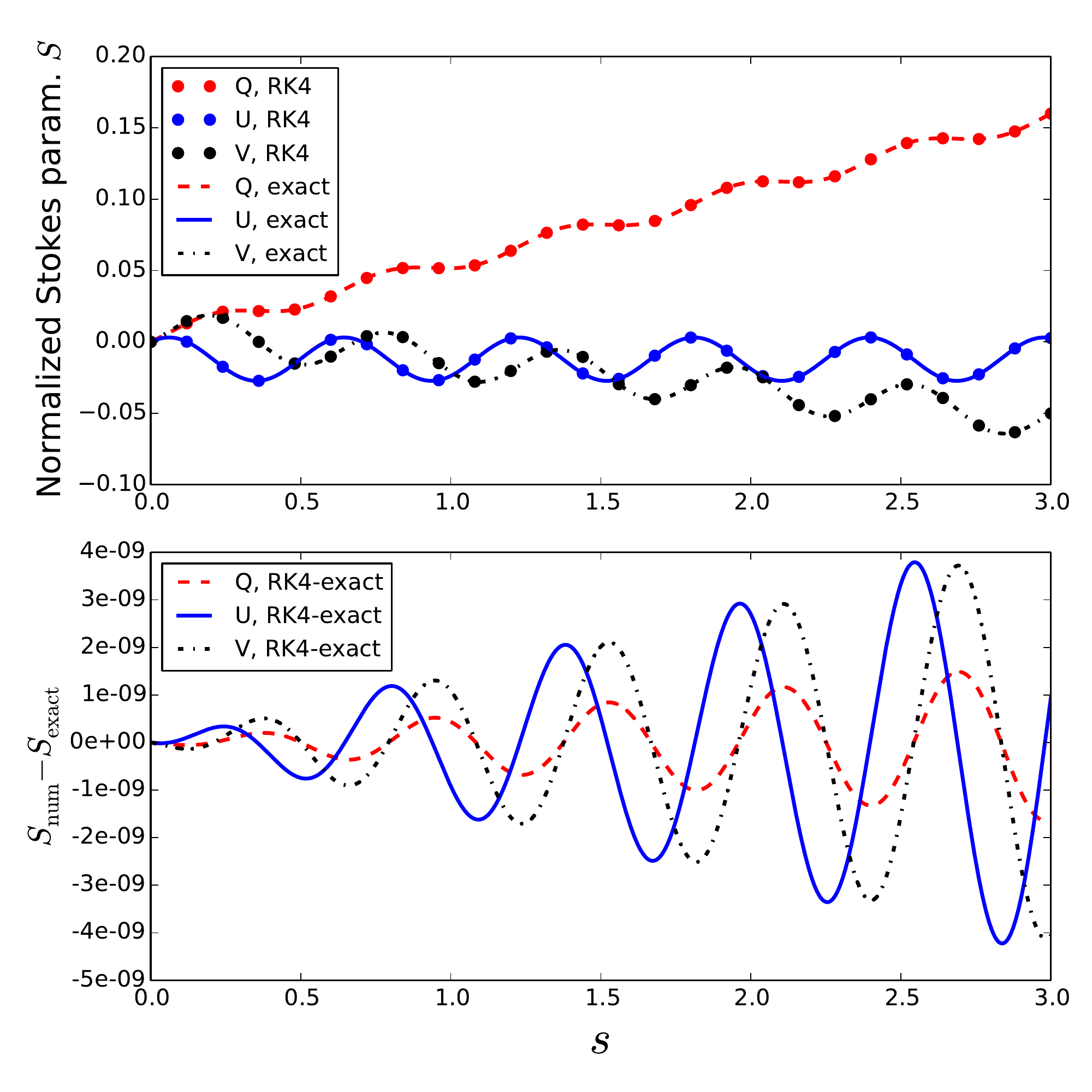}
\caption{A plot of Stokes Q, U, and V (in normalized units) as a function of distance traveled $s$, using the RK4 integration routine, for the second flat-spacetime plasma-integration test.}\label{fig:plasmatest2}
\end{figure}

\begin{figure}
\centering
\includegraphics[width=0.5 \textwidth]{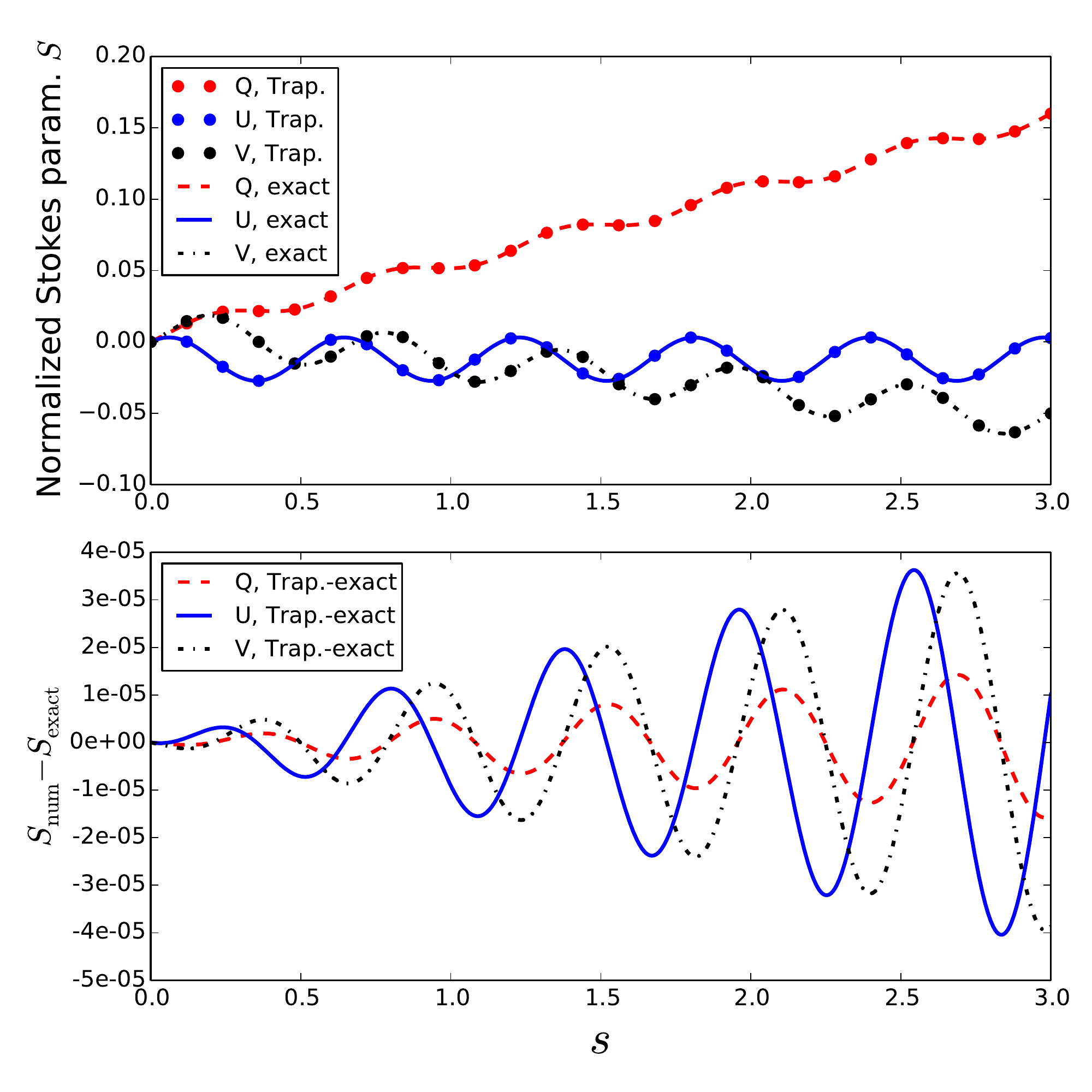}
\caption{A plot of Stokes Q, U, and V (in normalized units) as a function of distance traveled $s$, using the implicit trapezoid integration routine, for the second flat-spacetime plasma-integration test.}\label{fig:plasmatest2_IE}
\end{figure}

\begin{figure}
\centering
\includegraphics[width=0.5 \textwidth]{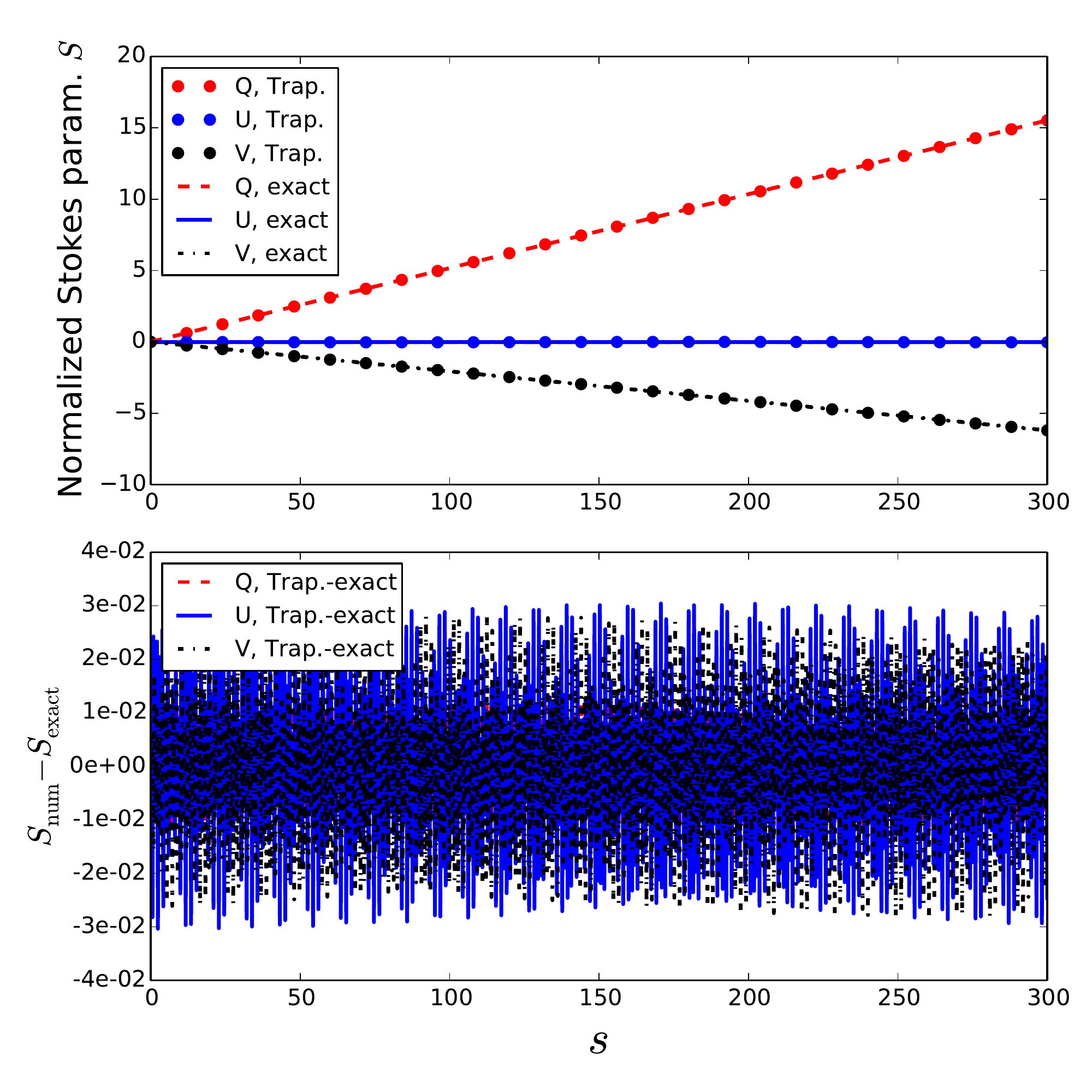}
\caption{As Fig.~\ref{fig:plasmatest2_IE}, but with a stepsize that is 100 times bigger, so the range of $s$ is 100 times larger as well. Note that the implicit trapezoid algorithm remains stable; the RK4 algorithm breaks down (the error diverges) under these conditions.}\label{fig:plasmatest2_bigstep}
\end{figure}

In the first plasma test, $j_I=2$, $j_Q=1$, $\alpha_I=1$, and $\alpha_Q=1.2$ (all other coefficients vanish). Figure~\ref{fig:plasmatest1} shows the integration results for this test using the RK4 algorithm, while Fig.~\ref{fig:plasmatest1_IE} shows the results for the implicit trapezoid algorithm. In the second plasma test, $j_Q=j_U=j_V=0.1$, $\rho_Q=10$, and $\rho_V=-4$ (again, all other coefficients vanish). Figure~\ref{fig:plasmatest2} shows the integration results for this test using the RK4 algorithm, while Fig.~\ref{fig:plasmatest2_IE} shows the results for the implicit trapezoid algorithm. 
%The results for the first test, using both integrators, show that the error of the numerical method grows rapidly at first, and then decays again. In the second test, the error grows linearly with the integration parameter $s$. 
Note the difference in scale of the errors for the implicit trapezoid scheme versus the explicit RK4 scheme - the RK4 scheme, being of a higher order, produces a more accurate result in both cases. On the other hand, the implicit trapezoid scheme is much more robust with respect to increasing the step size, as is shown in Fig.~\ref{fig:plasmatest2_bigstep}, which is a repeat of Fig.~\ref{fig:plasmatest2_IE}, but with a stepsize that is 100 times larger than previously. Using these settings, the RK4 scheme fails to produce a meaningful result, and the error diverges; the trapezoid scheme, on the other hand, remains stable, although the accuracy is affected by the large stepsize.

%As was explained in Section~\ref{sec:plasma_integration}, the potential weaknesses of both the implicit and explicit routines are illustrated in Figs.~\ref{fig:RK4_explodes} and \ref{fig:IE_dampens}. 

\begin{figure}
\centering
\includegraphics[width=0.5 \textwidth]{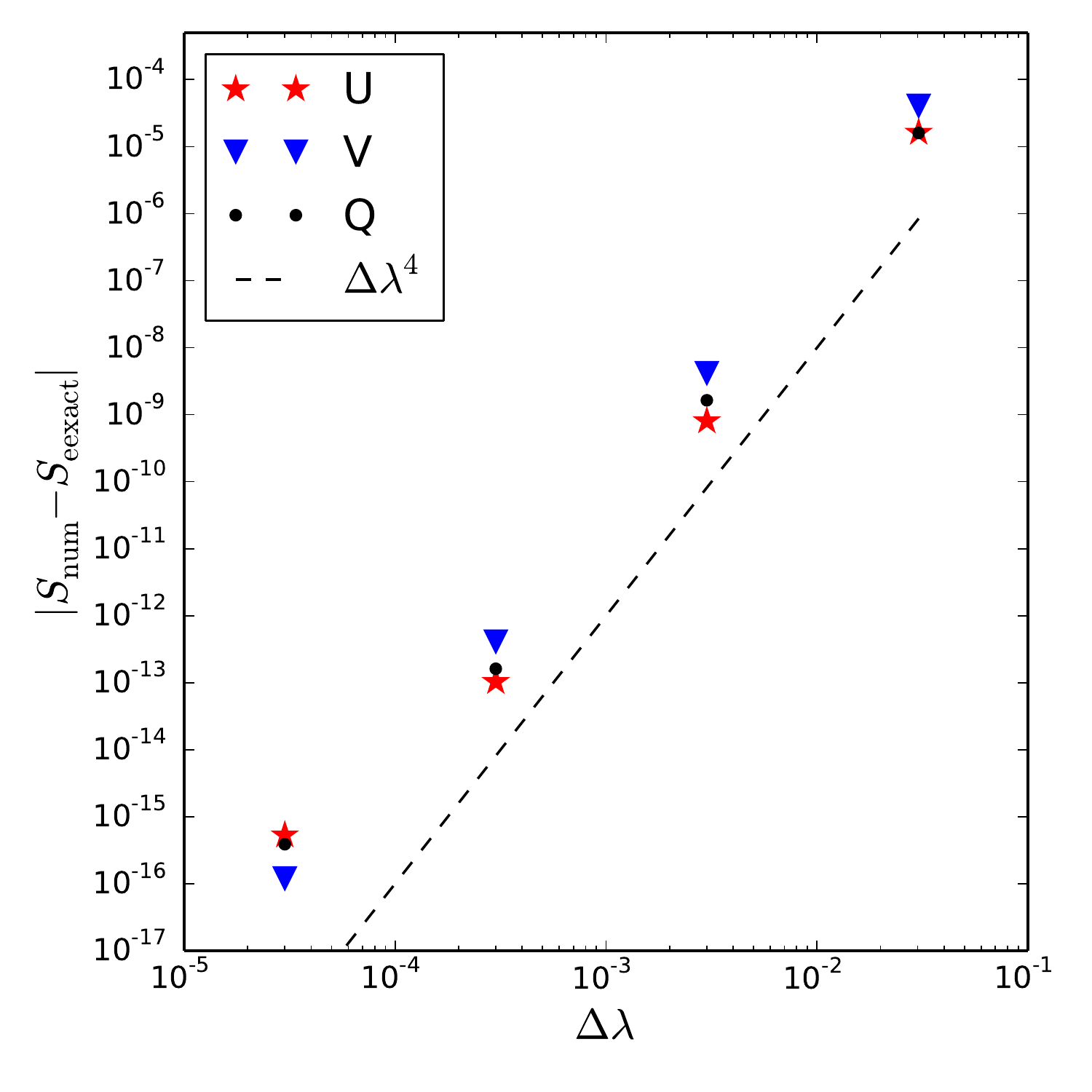}
\caption{Stepsize-convergence plot for the 4th-order Runge-Kutta (RK4) plasma-integration routine, for the second plasma test (see Section~\ref{sec:plasmatests}). The error is proportional to $\Delta \lambda^{4}$, as is expected. The fact that the leftmost datapoint is slightly offset from the convergence line suggests that the machine precision, which is of order $1{\rm e}{-16}$ for the double-precision arithmetic used in {\tt RAPTOR}, is reached.}\label{fig:plasma_RK4_convergence}
\end{figure}

\begin{figure}
\centering
\includegraphics[width=0.5 \textwidth]{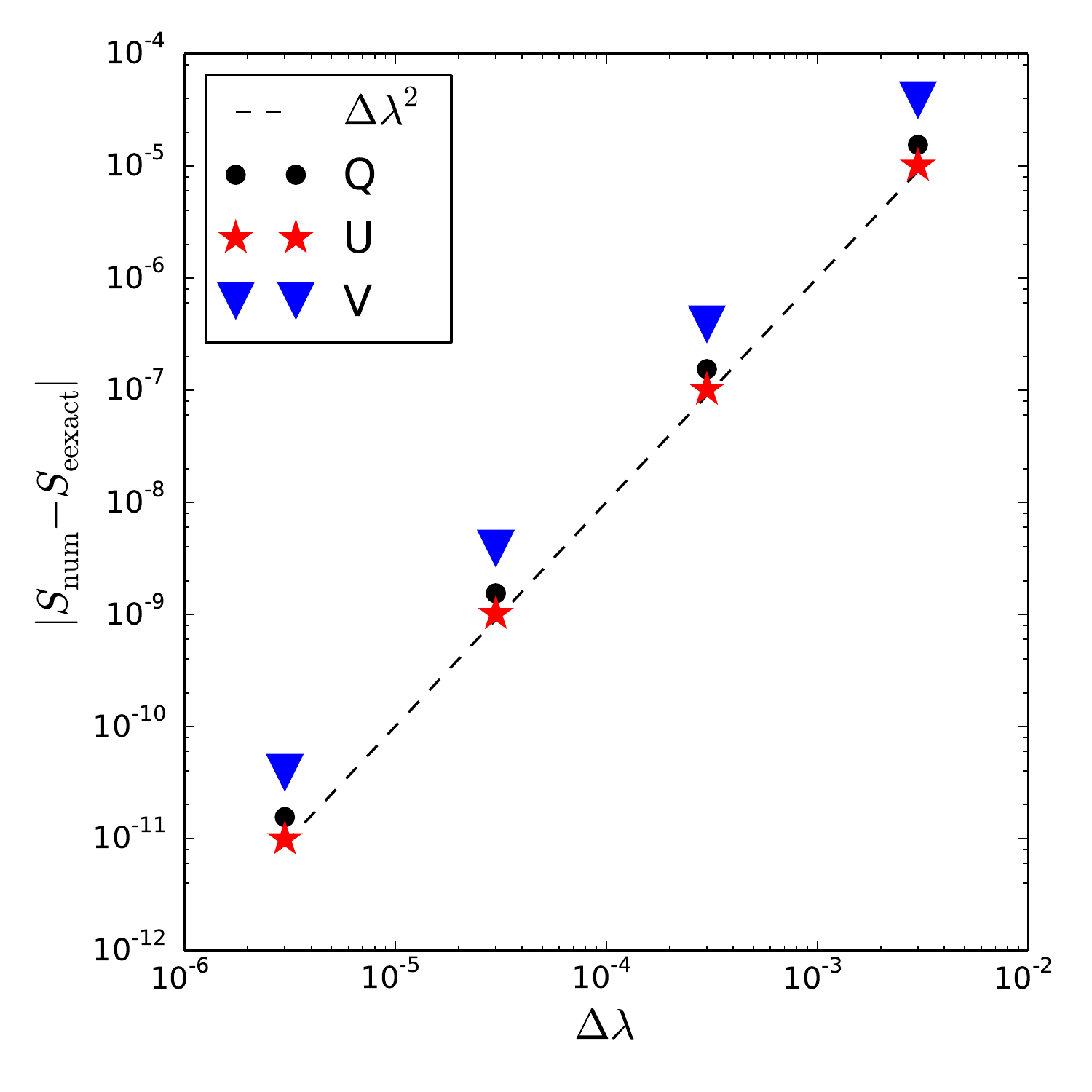}
\caption{Stepsize-convergence plot for the implicit trapezoid (IT) plasma-integration routine, for the second plasma test (see Section~\ref{sec:plasmatests}). The error is proportional to $\Delta \lambda^2$, as is expected.}\label{fig:plasma_IE_convergence}
\end{figure}

Error-convergence plots for both routines are shown in Figs.~\ref{fig:plasma_RK4_convergence} and \ref{fig:plasma_IE_convergence} (to compute the error for these plots, the absolute difference between the exact and the numerical solution was taken at the end of integration, where $\lambda=3$.)

%\ZY{Discussion about the implicit vs explicit methods here too.}
%\textcolor{red}{Note that this test does not trigger the stiff integrator, it is easy enough to use RK4 all the way through.}

%\begin{figure}
%\centering
%\includegraphics[width=0.5 \textwidth]{Figures/RK4_explodes.pdf}
%\caption{An illustration of the weakness of explicit methods for a stiff problem: a modified version of the second plasma test. Here, the problem has been made more stiff, by setting $\rho_Q=100$ and $\rho_V=-40$. The numerical solution initially tracks the exact one, but it quickly oscillates out of control.}\label{fig:RK4_explodes}
%\end{figure}

%\begin{figure}
%\centering
%\includegraphics[width=0.5 \textwidth]{Figures/IE_dampens.pdf}
%\caption{An illustration of the weakness of the implicit trapezoid method: the dampening of high-frequency oscillations. For this figure, the stepsize has been increased with respect to Fig.~\ref{fig:plasmatest2_IE}, exaggerating the problem. Note that the numerical solution tracks the oscillations initially, but then it quickly dampens out to a set of straight lines, which follow only the basic contours of the exact solution.}\label{fig:IE_dampens}
%\end{figure}

\subsection{Spacetime-integration test; thin-disk model}
\label{sec:thindisk}

\begin{figure*}
\centering
\includegraphics[width=0.99 \textwidth]{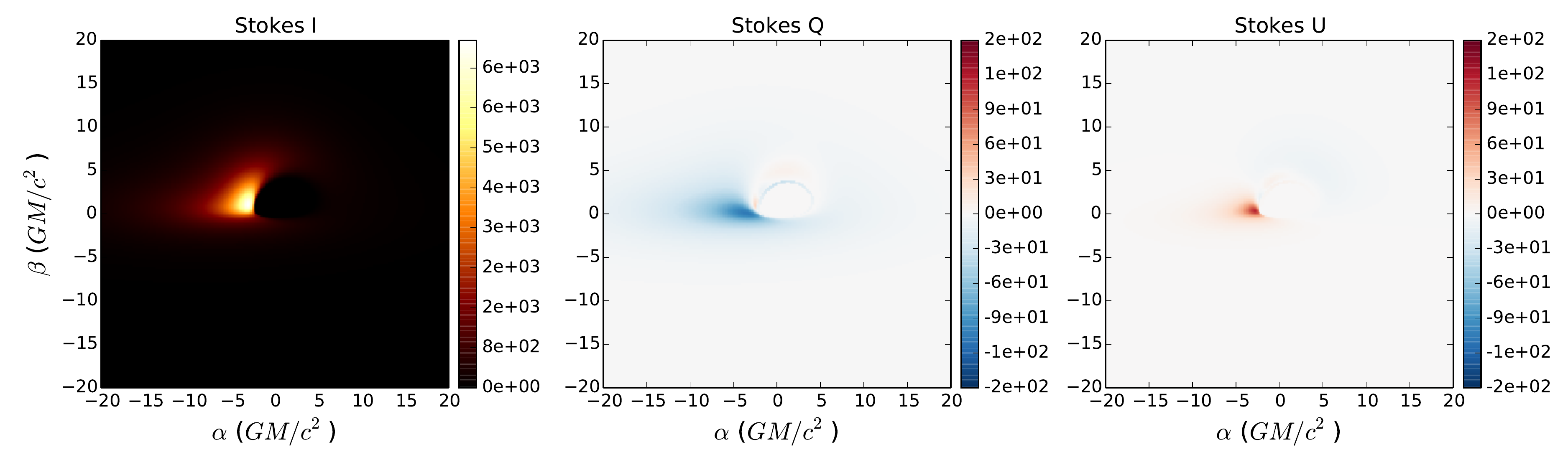}
\caption{Polarized image of the thin-disk model from \citet{schnittman2009} in terms of the Stokes parameters. The image size is 200 by 200 pixels. Flux is given in $\text{Jy} \ \text{px}^{-1}$ (Jansky per pixel). Stokes V is omitted, as it vanishes due to the symmetry of the problem.}\label{fig:thindiskIQUV}
\end{figure*}

\begin{figure}
\centering
\includegraphics[width=0.5 \textwidth]{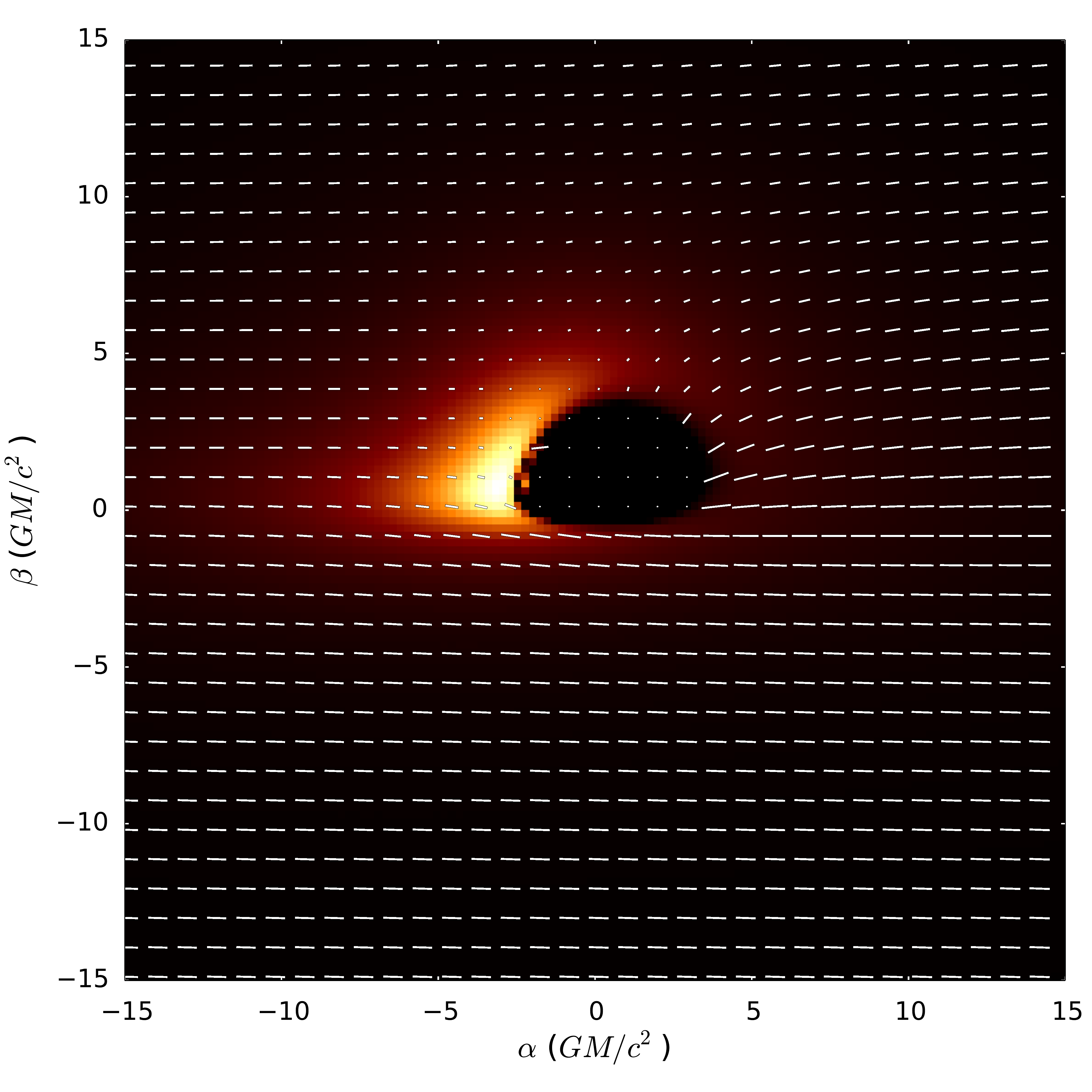}
\caption{Polarized image of the thin-disk model from \citet{schnittman2009} shown with polarization vectors. }\label{fig:thindisk}
\end{figure}

Next, we test our integrator for Eq.~\ref{eqn:spacetime_propagation}, i.e., the equation for propagation of a polarized ray through spacetime. We do so in a spacetime devoid of matter, save for a geometrically thin, optically thick accretion disk in the equatorial plane, based on the description by \citet{shakura1973} and \citet{novikovthorne1973}. The blackbody radiation emitted by the disk is scattered by electrons in the disk's atmosphere. The radiation is limb-darkened by the disk's atmosphere (note that the atmosphere is not represented in the geometry of the model, but its effects are taken into account by modifying the emission coefficient), and it is partially linearly polarized in the plane of the disk, perpendicular to both the wave vector and the disk normal, as in the model presented in \citet{chandrasekhar1960} and studied in \citet{connors1980} and \citet{schnittman2009}. In this model, the emission coefficients are delta functions along the ray; they are evaluated once, after which the polarized light is transported through the vacuum, allowing us to test the vacuum-integration routine (this result was earlier reproduced in \citet{dexter2016}).

Figure~\ref{fig:thindiskIQUV} shows the results obtained using the settings specified by an EHT internal note on polarized radiative transfer; these settings are listed in Table~\ref{tab:test_settings}. The integrated flux densities for all Stokes parameters are reported in Table~\ref{tab:thindisk_fluxes}. Note that, although we report the results in units of Jansky per pixel for continuity with the rest of the paper, the camera frequency for this result was in the X-ray part of the electromagnetic spectrum. Figure~\ref{fig:thindisk} shows the same result but, in the `polarization vector representation', to be compared with Fig.~1 in \citet{schnittman2009}. Figure~\ref{fig:thindisk} also uses a value of $0.9$ for the black-hole spin, as did those authors.

\begin{table}[h]
\bgroup
\def\arraystretch{1.25}
\begin{tabularx}{0.49\textwidth}{@{}p{0.075\textwidth} p{0.11\textwidth} p{0.12\textwidth} p{0.12\textwidth}@{}}
      \thickhline
      \textbf{Flux} & \textbf{Thin disk} & \textbf{GRMHD low} & \textbf{GRMHD high}\\
      \hline
$S_{\nu,I} \left(\text{Jy}\right)$ & $6.869 \cdot 10^{6}$ & $0.0102$ & $0.494$ \\ 
$S_{\nu,Q} \left(\text{Jy}\right)$ & $-1.586 \cdot 10^{5}$  & $-0.000312$  & $-0.00722$  \\ 
$S_{\nu,U} \left(\text{Jy}\right)$ & $1.057 \cdot 10^{4}$  & $0.00032$  & $-0.00328$  \\ 
$S_{\nu,V} \left(\text{Jy}\right)$ & $0$                            & $-6.96 \cdot 10^{-6}$  & $0.00333$  \\
      \thickhline
    \end{tabularx}
\egroup
\caption{Integrated flux densities for all Stokes parameters for the thin-disk test (Section \ref{sec:thindisk}) as well as the low- and high-flux versions of the GRMHD test (Section \ref{sec:plasmatests}). Images of these tests (albeit at a modified field of view and camera resolution) are shown in Figs.~\ref{fig:thindiskIQUV}, \ref{fig:test3low}, and \ref{fig:test3high}.}
\label{tab:thindisk_fluxes}
\end{table}

\begin{table}[h]
\bgroup
\def\arraystretch{1.25}
\begin{tabularx}{0.49\textwidth}{@{}p{0.1\textwidth} p{0.19\textwidth} p{0.2\textwidth}@{}}
      \thickhline
      \textbf{Parameter} & \textbf{Thin-disk test value} & \textbf{GRMHD test value}\\
      \hline
$a$ & $0.99$ & $0.94$ \\
$\text{M}_{\text{BH}}$ & $10 \msun$ & $6.2 \cdot 10^9  \msun$ \\
$d_{\text{source}}$ & $0.05$ pc & $16.9$ Mpc \\
$\nu_{cam}$ & $2.417989 \cdot 10^{17}$ Hz & 230 GHz \\
$r_{cam}$ &  $10^4 R_G$ & $10^4 R_G$ \\
$\theta_{cam}$ & $75 \deg$ & $163 \deg$ \\
$\phi_{cam}$ & 0 & 0 \\
$DX=DY$ & $40$ $R_G$ & $44.17 R_G$ \\
$NX=NY$ & $80$ px & $160$ px \\
$\dot{\text{M}}$ & $1.399 \cdot 10^{17} {\rm g/s}$ & $-$ \\
$\mathcal{M}_{\text{low}}$ & $-$ & $1.672 \cdot 10^{26}$ g \\
$\mathcal{M}_{\text{high}}$ & $-$ & $2.739 \cdot 10^{25}$ g \\
      \thickhline
    \end{tabularx}
\egroup
\caption{Settings used to reproduce the thin-disk test (Section \ref{sec:thindisk}) as well as the low- and high-flux versions of the GRMHD test (Section \ref{sec:plasmatests}). The test results are recapitulated in Table \ref{tab:thindisk_fluxes}.}
\label{tab:test_settings}
\end{table}

\begin{figure}
\centering
\includegraphics[width=0.5 \textwidth]{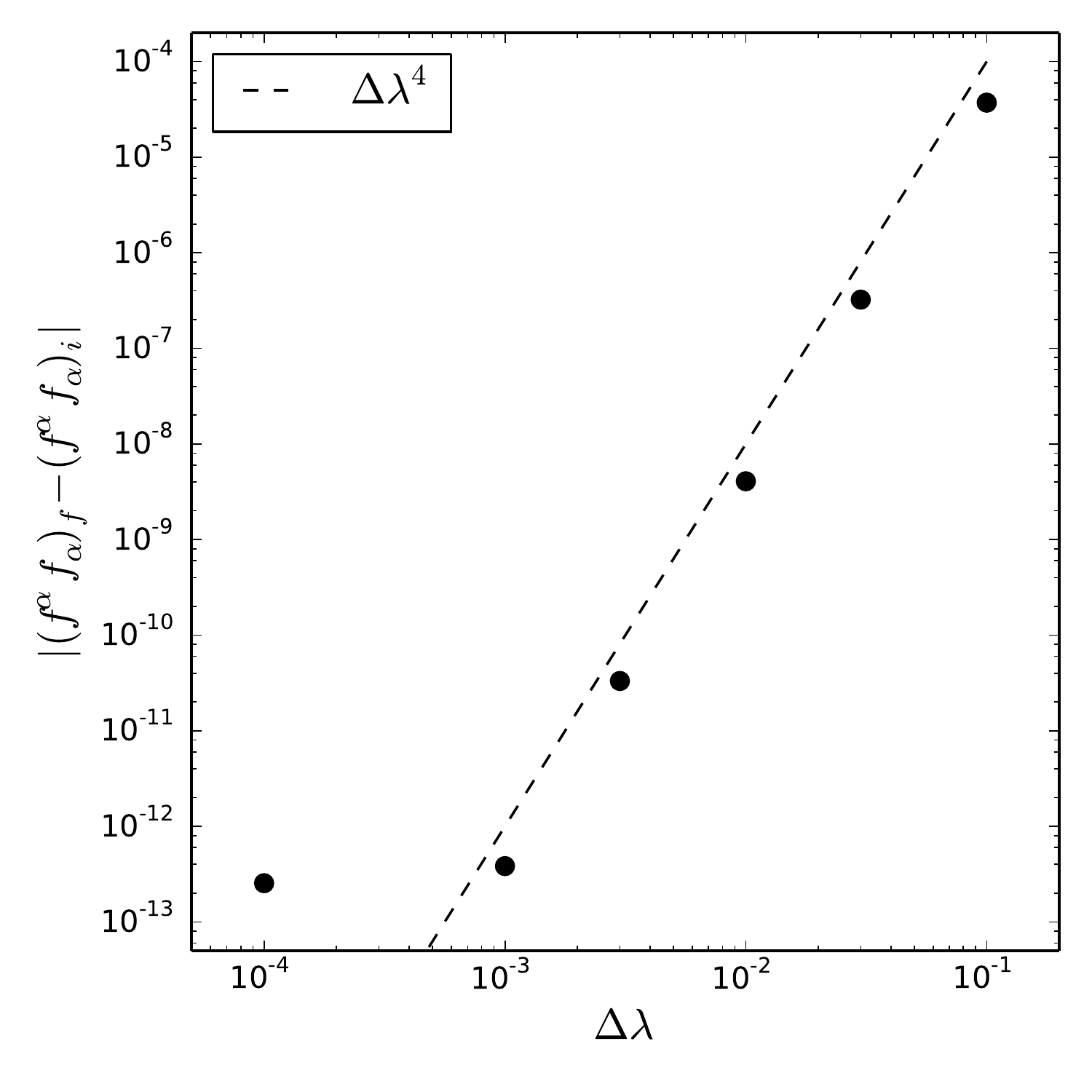}
\caption{Stepsize-convergence plot for the 4th-order Runge-Kutta (RK4) spacetime-transport routine, for a particular ray in the thin-disk test (see Section \ref{sec:thindisk}). The error is proportional to $\Delta \lambda^{4}$, as is expected.}\label{fig:transport_RK4_convergence}
\end{figure}

Figure~\ref{fig:transport_RK4_convergence} shows the stepsize-convergence plot for the RK4 spacetime-integration routine (which integrates Eq.~\ref{eqn:spacetime_propagation}). The error was computed by tracking the norm of the polarization vector, which should be conserved after integration of the ray, which passes close to the black hole, through a strongly curved region of spacetime.

\subsection{Spacetime-and-plasma (GRMHD) integration test}
\label{sec:grmhd_test}

\begin{figure*}
\centering
\includegraphics[width= \textwidth]{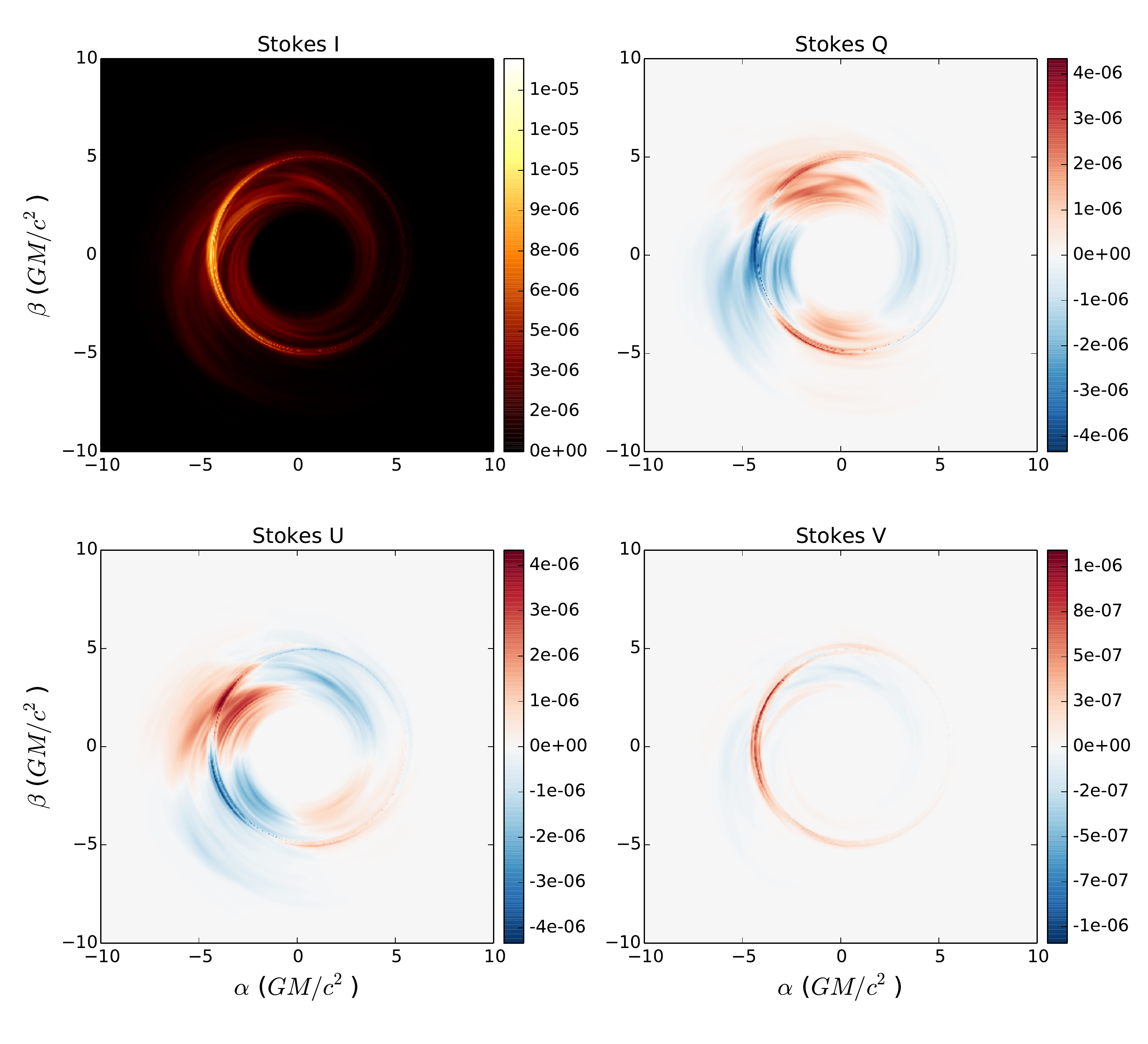}
\caption{Polarized image of the GRMHD-based third test for the high-accretion-rate scenario. In order to enhance the clarity of these images, we have chosen a different field of view from the original test (see Table \ref{tab:test_settings}); the camera size for these images is 20 by 20 $R_G$, and the image size is 1024 by 1024 pixels. Flux is given in $\text{Jy} \ \text{px}^{-1}$ (Jansky per pixel). The integrated flux densities, useful for code-comparison purposes, are recapitulated in Table \ref{tab:thindisk_fluxes}.}\label{fig:test3low}
\end{figure*}

\begin{figure*}
\centering
\includegraphics[width= \textwidth]{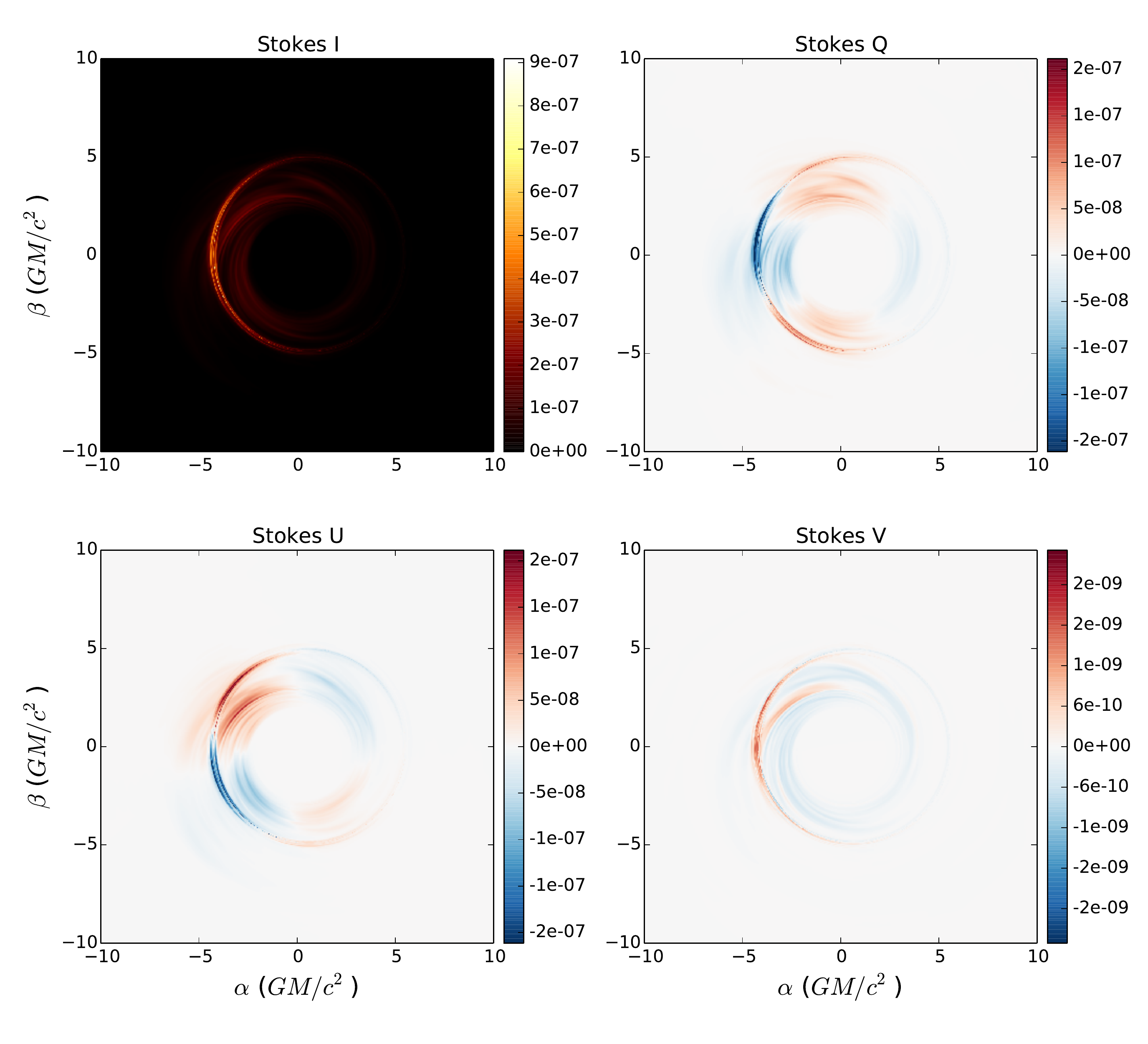}
\caption{Polarized image of the GRMHD-based third test for the low-accretion-rate scenario. In order to enhance the clarity of these images, we have chosen a different field of view from the original test (see Table \ref{tab:test_settings}); the camera size for these images is 20 by 20 $R_G$, and the image size is 1024 by 1024 pixels. Flux is given in $\text{Jy} \ \text{px}^{-1}$ (Jansky per pixel). The integrated flux densities, useful for code-comparison purposes, are recapitulated in Table \ref{tab:thindisk_fluxes}.}\label{fig:test3high}
\end{figure*}

Finally, we test the combination of the vacuum and plasma-integration routines. To do so, the rays must propagate through a radiating plasma that is of finite extent (i.e., not infinitesimally thin), so that plasma interaction takes place as the rays are propagating through curved spacetime. This may be achieved using data from a GRMHD simulation of an accreting AGN; a particular data dump, created using the {\tt HARM} code \citep{gammie2003}, was used for these tests, which are part of the EHT's internal code-comparison efforts for polarized radiative transfer, a forthcoming publication. The radiative model uses the thermal-synchrotron emission, absorption, and rotation coefficients listed in \ref{app:coefficients}, uses a constant proton-to-electron temperature ratio of 3, making for a disk-dominated emission model. To reproduce the results listed in this paper, please refer to Table \ref{tab:test_settings}, which shows the {\tt RAPTOR} settings used to generate the test results for two scenarios (the low-accretion-rate and high-accretion-rate scenarios, respectively); the GRMHD dump that was used to generate these results is distributed along with {\tt RAPTOR}. The resulting images (Figs.~\ref{fig:test3low} and \ref{fig:test3high}) show a low-luminosity AGN at a low inclination angle (163 degrees, or 17 degrees from the southern pole). A radial pattern is observed. Table \ref{tab:thindisk_fluxes} lists the flux densities obtained for the thin-disk and GRMHD tests.

\subsection{Convergence with other polarized radiative-transfer codes}

The thin-disk and GRMHD tests discussed in the previous two sections have been adopted by the EHT's collaborative effort to establish convergence between polarized general-relativistic radiative transfer codes, which is to be published in a forthcoming paper. At the time of publication, comparison to {\tt ipole} output for the same tests yielded an agreement to order 1\% in terms of mean squared error. Experience has revealed a few possible sources of error, such as the precise camera setup used, or the approximations used to compute Bessel functions in the emission/absorption/rotation coefficients. However, since there are uncertainties that are larger than 1\% in such aspects of the computation as the microphysics (e.g., the energy-distribution functions of the radiating electrons, which is affected by magnetic reconnection, fine-scale turbulence and shockwaves, etc.), the agreement that has been achieved is sufficient for current EHT science goals. Nevertheless, these results may be refined further, and are to be published in a forthcoming EHT paper comparing methods for polarized general-relativistic radiative transfer.

%\section{Results}
%\label{sec:results}

%These are our results.

%\section{Scientific application?}

%If possible we should apply the code to an astrophysics problem here...

\section{Summary}
\label{sec:discussion}

In order to deepen understanding of polarized observations of astrophysical phenomena originating in strong gravitational fields, we have developed a code capable of performing efficient polarized radiative transfer in arbitrary spacetimes. Our implementation can represent a polarized ray using both the Stokes vector and the polarization vector (plus relevant intensities), and can switch between the two. The former description is appropriate for plasma interactions (whenever the ray is inside a plasma), while the latter presents a better description for spacetime propagation. This formalism is conceptually simple and uses the minimum number of degrees of freedom.

The polarized radiative-transfer equation, which describes a ray's interaction with a plasma, may become stiff in some regions of the integration volume, while being much easier to integrate in others. We have developed a computationally efficient yet robust implicit-explicit scheme to integrate this equation. In order to maximize accuracy and efficiency, our algorithm switches between implicit and explicit integration schemes. We have determined when the polarized radiative-transfer equation (as expressed in our particular tetrad frame) becomes stiff for the fourth-order Runge-Kutta integrator in order to establish the switching criterium.

We have demonstrated correctness of our new algorithms for spacetime propagation and plasma interaction separately as well as together. Comparison of {\tt RAPTOR} output for the thin-disk and GRMHD tests described in Sections \ref{sec:thindisk} and \ref{sec:grmhd_test} to that of {\tt ipole} show excellent agreement, strongly suggesting that the two different implementations agree on the physics to a sufficient degree of accuracy for the Event Horizon Telescope, in the context of a relevant, complex astrophysical problem. The value in that result lies both in adding credence to the EHT's theoretical calculations pertaining to polarized sources such as M87, and in providing a new, public tool for the efficient production of simulated, polarized observations.

Polarized observations have the potential to allow observers to determine the structure of magnetic fields in the accretion disks and jets of black holes. Meanwhile, theoretical investigations probe the observational effects produced by different plasma models (see, e.g., \citealt{rosales2018}; \citealt{palumbo2020}), and even different theories of gravity \citep{mizuno2018bb}. RAPTOR can be used as an efficient and flexible tool with which to explore the radiative properties of different plasma models in arbitrary spacetimes. Such simulations are to be compared with current and future observations of polarized radiation emitted by relativistic plasma's orbiting black holes, neutron stars, and potentially other, more exotic objects.

\section*{Acknowledgements}

This work is supported by the ERC Synergy Grant "BlackHoleCam: Imaging the Event Horizon of Black Holes" (Grant 610058).
ZY is supported by a Leverhulme Trust Early Career Fellowship. The authors thank Koen de Boer, Monika Mo{\'s}cibrodzka, Ben Prather, George Wong, and the anonymous referee for their helpful suggestions regarding the manuscript. This work has made use of NASA's Astrophysics Data System (ADS). 

\newpage

\bibliography{RAPTOR_II_AA}
\bibliographystyle{apalike}

%
%-------------------------------------------------------------
%          For the appendices, table longer than a single page
%-------------------------------------------------------------

% Table will be print automatically at the end of the document, 
% after the whole appendices

\newpage 

\begin{appendix} 

\section{Stiffness analysis of the polarized radiative-transfer equation}
\label{sec:stiffness}

The stiffness of a linear system of differential equations, such as the polarized radiative-transfer equation (Eq.~\ref{eqn:plasma_interaction}), with respect to a particular explicit integration scheme (such as the fourth-order Runge-Kutta integrator used here), depends on the eigenvalues of the matrix which appears on that equation's right-hand side, ${\bf M}$; our first step is to compute them. 

In a frame in which $j_U = \alpha_U = \rho_U = 0$, ${\bf M}$'s characteristic polynomial is given by
\begin{equation}
\left| {\bf M} \right| = z^2 + a_2 z + a_0 = 0,    
\end{equation}
where
\begin{subequations}
\begin{align}
    z &= \left( \alpha_I - \Lambda \right)^2, \\
    a_2 &= \rho_Q^2 + \rho_V^2 - \alpha_Q^2 - \alpha_V^2,\\
    a_0 &= -2 \alpha_Q \alpha_V \rho_Q \rho_V - \alpha_Q^2 \rho_Q^2 - \alpha_V^2 \rho_V^2,
\end{align}
\end{subequations}
$\Lambda$ being an eigenvalue of ${\bf M}$. In other words, ${\bf M}$'s characteristic polynomial is a biquadratic equation in $\left( \alpha_I - \Lambda \right)$, which may therefore be obtained by solving the quadratic equation:
\begin{equation}
z_{\pm} = \frac{-a_2 \pm \sqrt{a_2^2 - 4 a_0}}{2}.
\end{equation}
There are then four possible values for $\left(\alpha_I - \Lambda \right)$:
\begin{subequations}
\begin{align}
      \left( \alpha_I - \Lambda \right) &=
  \begin{cases}
    &+\sqrt{z_{+}}, \\
    &-\sqrt{z_{+}}, \\
    &+\sqrt{z_{-}}, \\
    &-\sqrt{z_{-}}, 
  \end{cases}
\end{align}
\end{subequations}
so that
\begin{equation}
    \Lambda = \alpha_I \pm \sqrt{z_{\pm}},
\end{equation}
where either plus/minus symbol may be interpreted in either way.

Now that ${\bf M}$'s eigenvalues are known, the stiffness of Eq.~\ref{eqn:plasma_interaction} for the explicit fourth-order Runge-Kutta integrator can be computed. Defining
\begin{equation}
    \zeta = \Delta \lambda \Lambda,
\end{equation}
where $\Delta \lambda$ is {\tt RAPTOR}'s integration step size, the explicit fourth-order Runge-Kutta integration routine is stable when
\begin{equation}
    \left| 1 + \zeta + \frac{1}{2} \zeta^2 + \frac{1}{6} \zeta^3 + \frac{1}{24} \zeta^4 \right| < 1.
    \label{eq:stability}
\end{equation}
Whenever this condition is not met for any one of ${\bf M}$'s four eigenvalues, it is necessary to switch to the implicit trapezoid integration routine. 

\newpage

\section{Implicit trapezoid integrator}
\label{app:implicit_euler}

Rewriting Eq.~\ref{eqn:implicit_euler}, we obtain a system of equations for $\mathbfcal{S}_{new}$:
\begin{equation}
    \underbrace{\left( \mathds{1} + \frac{\Delta \lambda}{2} \text{\bf M} \right)}_\text{\bf A}  \mathbfcal{S}_{new} = \underbrace{\mathbfcal{S} + \frac{\Delta \lambda}{2} \Bigg( 2 {\bf j} - \text{\bf M} \mathbfcal{S} \Bigg)}_\text{\bf b}.
\end{equation}
The matrix $\text{\bf A}$, evaluated in a frame in which $j_U$, $\alpha_U$, and $\rho_U$ vanish, is then given by
\begin{equation}
\text{\bf A} =
         \begin{pmatrix}
           1 + \Delta \lambda \alpha_I  / 2 & \Delta \lambda \alpha_Q  / 2 & 0 & \Delta \lambda \alpha_V  / 2 \\
           \Delta \lambda \alpha_Q  / 2 & 1 + \Delta \lambda \alpha_I  / 2 & \Delta \lambda \rho_V  / 2 & 0  \\
           0 & -\Delta \lambda \rho_V  / 2 & 1 + \Delta \lambda \alpha_I / 2  & \Delta \lambda \rho_Q  / 2  \\
           \Delta \lambda \alpha_V  / 2 & 0   & -\Delta \lambda \rho_Q  / 2 & 1 + \Delta \lambda \alpha_I  / 2
         \end{pmatrix}.
\end{equation}
Next, we express $\text{\bf A}$ as a product of two triangular matrixes, $\text{\bf L}$ and $\text{\bf U}$:
\begin{equation}
\text{\bf A}  = \text{\bf L} \text{\bf U} =
         \begin{pmatrix}
           1      & 0      & 0   & 0  \\
           l_{21} & 1      & 0   & 0  \\
           l_{31} & l_{32} & 1   & 0  \\
           l_{41} & l_{42} & l_{43} & 1 
         \end{pmatrix}
         \begin{pmatrix}
           u_{11} & u_{12} & u_{13} & u_{14} \\
           0 & u_{22}      & u_{23}   & u_{24}  \\
           0 & 0 & u_{33}   & u_{34}  \\
           0 & 0 & 0 & u_{44} 
         \end{pmatrix},
\end{equation}
whose elements are given by
\begin{subequations}
\begin{align}
u_{11} &= 1 + \Delta \lambda \alpha_I  / 2, \\
u_{12} &= \Delta \lambda \alpha_Q  / 2, \\
u_{13} &= 0, \\
u_{14} &= \Delta \lambda \alpha_V  / 2, \\
l_{21} &= \frac{\Delta \lambda \alpha_Q}{2 u_{11}}, \\
u_{22} &= 1 + \Delta \lambda \alpha_I  / 2 - l_{21} u_{12}, \\
u_{23} &= \Delta \lambda \rho_V  / 2, \\
u_{24} &= - l_{21} u_{14}, \\
l_{31} &= 0, \\
l_{32} &= -\frac{\Delta \lambda \rho_V}{2 u_{22}}, \\
u_{33} &= 1 + \Delta \lambda \alpha_I / 2 - l_{32} u_{23}, \\
u_{34} &= \Delta \lambda \rho_Q  / 2 - l_{32} u_{24}, \\
l_{41} &= \frac{\Delta \lambda \alpha_V}{2 u_11}, \\
l_{42} &= -\frac{l_{41} u_{12}}{u_{22}}, \\
l_{43} &= -\frac{\Delta \lambda \rho_Q  / 2 + l_{42} u_{23}}{u_{33}}, \\
u_{44} &= 1 + \Delta \lambda \alpha_I  / 2 - l_{41} u_{14} - l_{42} u_{24} - l_{43} u_{34}.
\end{align}
\label{eqn:LUcoeffs}
\end{subequations}
This allows us to obtain $\mathbfcal{S}_{new}$ explicitly, from two linear systems of equations:
\begin{subequations}
    \begin{align}
        &\text{\bf L} \text{\bf y} = \text{\bf b}, \\
        &\text{\bf U} \mathbfcal{S}_{new} = \text{\bf y},
    \end{align}
\end{subequations}
where $\text{\bf y}$'s components are given by
\begin{subequations}
\begin{align}
    y_1 &= b_1, \\
    y_2 &= b_2 - l_{21} y_1, \\
    y_3 &= b_3 - l_{32} y_2, \\
    y_4 &= b_4 - \left( l_{41} y_1 + l_{42} y_2 + l_{43} y_3 \right).
\end{align}
\end{subequations}
$\mathbfcal{S}_{new}$ is then computed as follows:
\begin{subequations}
\begin{align}
    \mathcal{S}_{new,1} &= \frac{y_1 - u_{12}x_2 - u_{14}x_4}{u_{11}}, \\
    \mathcal{S}_{new,2} &= \frac{y_2 - u_{23} x_3 - u_{24}x_4}{u_{22}}, \\
    \mathcal{S}_{new,3} &= \frac{y_3-u_{34}x_4}{u_{33}}, \\
    \mathcal{S}_{new,4} &= \frac{y_4}{u_{44}}.
\end{align}
\end{subequations}

\newpage

\section{Emission, absorption, and rotation coefficients for a thermal electron population}
\label{app:coefficients}

This appendix lists the emission, absorption, and rotation coefficients employed in {\tt RAPTOR}. They pertain to a (relativistic) thermal distribution of electrons. They are adapted from \citet{dexter2016} (note that non-thermal (power-law) coefficients are also listed in that paper), with a number of modifications of numerical fit functions by \citet{ipole}, as well as minor notational rewrites in eqs.~\ref{eq:rots} and typographical corrections in eq.~\ref{eq:fmod} (cf. B14 in \citet{dexter2016}).

The emission coefficients are given by
\begin{subequations}
\begin{align}
j_I &= \frac{n_e e^2 \nu}{2 \sqrt{3} c {\theta_e}^2} I_I\left( x \right), \\
j_Q &= \frac{n_e e^2 \nu}{2 \sqrt{3} c {\theta_e}^2} I_Q\left( x \right), \\
j_V &= \frac{2 n_e e^2 \nu }{ 3 \sqrt{3} c {\theta_e}^3 \tan{\theta_B}} I_V\left( x \right),
\end{align}
\end{subequations}
where $n_e$ is the electron number density, $e$ is the electron charge, $c$ is the speed of light, $\theta_e$ is the dimensionless electron temperature, $\theta_B$ is the angle between the wave vector and the magnetic-field vector, and $x$ is the ratio of the ray's frequency over the critical plasma frequency:
\begin{equation}
    x = \frac{\nu}{\nu_c},
\end{equation}
where
\begin{equation}
    \nu_c = \frac{3 e B \sin{\theta_B} \theta_e^2}{4 \pi m_e c},
\end{equation}
with $B$ being the magnetic-field amplitude (in Gaussian-cgs units).

The expressions $I_I, I_Q, I_V$ represent numerical fit functions, and are given by
\begin{subequations}
\begin{align}
I_I\left( x \right) &= 2.5651 \left(1 + 1.92 x^{-1/3} + 0.9977 x^{-2/3} \right) e^{-1.8899 x^{1/3}}, \\
I_Q\left( x \right) &= 2.5651 \left(1 + 0.93193 x^{-1 / 3} + 0.499873 x^{-2 / 3} \right) e^{-1.8899 x^{1 / 3}}, \\
I_V\left( x \right) &= (1.81348 / x + 3.42319 x^{-2 / 3} + 0.0292545 x^{-1/2} +\notag \\
&2.03773 x^{-1 / 3} ) e^{-1.8899 x^{1 / 3}}.
\end{align}
\end{subequations}
Absorption is computed using Kirchoff's law, as the distribution is thermal, so that the absorption coefficients may be written as
\begin{equation}
    \alpha_{\nu} = j_{\nu} / B_{\nu},
\end{equation}
where $B_{\nu}$ is the blackbody function.

Finally, the rotation coefficients are given by
\begin{subequations}
\begin{align}
\rho_Q &= \frac{ \omega_p^2 \omega_0^2 \sin^2{\theta_B}}{16 c \pi^3 \nu^3 } f_m\left( \tilde{X} \right) + \left[ \frac{K_1\left( \theta_e^{-1} \right)}{K_2\left(\theta_e^{-1} \right)} + 6 \theta_e \right], \\
\rho_V &= \frac{ \omega_p^2 \omega_0 \cos{\theta_B}}{ 4 c \pi^2 \nu^2} \frac{K_0 \left( \theta_e^{-1} \right) - \Delta J_5 \left( \tilde{X} \right)}{K_2\left( \theta_e^{-1} \right)},
\end{align}
\label{eq:rots}
\end{subequations}
where
\begin{equation}
\omega_p^2 = 4 \pi n_e e^2 / m_e,
\end{equation}
\begin{equation}
\tilde{X} = \left( \frac{3}{2 \sqrt{2}} 10^{-3} \frac{\nu}{\nu_c} \right)^{-1/2},
\end{equation}
and
\begin{equation}
\omega_0 = \frac{e B}{m_e c}.
\end{equation}
The functions $f_m$ and $\Delta J_5$ again represent fit functions. They are given by
\begin{multline*}
f_m\left(\tilde{X}\right) = 2.011 \exp{\left(-\frac{\tilde{X}^{1.035}}{4.7}\right)} - \cos{\left(\frac{\tilde{X}}{2}\right)} \exp{\left(-\frac{\tilde{X}^{1.2}}{2.73}\right)} -\\ 
0.011 \exp{\left(-\frac{\tilde{X}}{47.2} \right)} + \left( 0.011 \exp{\left( -\frac{\tilde{X}}{47.2} \right)} - \right. \\
\left. 2^{-1 / 3} 3^{-23 / 6} \pi 10^4 \tilde{X}^{-8 / 3} \right) \frac{1}{2} \left(1 + \frac{\tanh{\left(\log{\tilde{X}} - \log{120}\right)}}{0.1} \right)
\label{eq:fmod}
\end{multline*}
and
\begin{equation}
\Delta J_5 \left( \tilde{X} \right) = 0.4379 \log{\left( 1 + 0.001858 \tilde{X}^{1.503} \right)},
\end{equation}
respectively.

Note that during a {\tt RAPTOR} run, the Lorentz-invariant versions of these coefficients are employed. Transforming the emission, absorption, and rotation coefficients to their Lorentz-invariant forms proceeds as follows:
\begin{subequations}
\begin{align}
    j_{\rm inv} &= j / \nu^2, \\
    \alpha_{\rm inv} &= \alpha \nu, \\
    \rho_{\rm inv} &= \rho \nu.
\end{align}
\end{subequations}

\end{appendix}

\end{document}